\documentclass[12pt]{article}
\usepackage{putex}
\usepackage{graphicx}

\def\AY#1{}
\def\FB#1{}
\def\CH#1{}
\def\RR#1{}

\def\be{\begin{equation}}
\def\ee{\end{equation}}
\def\bea{\begin{equation}\begin{aligned}}
\def\eea{\end{aligned}\end{equation}}
\def\nn{\nonumber}
\def\mat#1{\begin{pmatrix} #1 \end{pmatrix}}

\newcommand{\du}[2]{_{#1}^{\phantom{#1}#2}}
\def\trans{{\ensuremath{\text{\sf T}}}}

\def\eg{{\textit{e.g.}}}

\newcommand{\cL}{\mathcal{L}}
\newcommand{\cM}{\mathcal{M}}

\newcommand{\cO}{\mathcal{O}}

\newcommand{\ulambda}{{\underline \lambda}}

\newcommand{\unu}{{\underline \nu}}

\newcommand{\usigma}{{\underline \sigma}}
\newcommand{\ut}{{\underline t}}
\newcommand{\ux}{{\underline x}}
\newcommand{\uy}{{\underline y}}
\newcommand{\uz}{{\underline z}}

\DeclareMathOperator{\Tr}{Tr}

\begin{document}

\preprint{PUPT-2342}

\institution{PU}{Department of Physics, Princeton University, Princeton, NJ 08544, USA}
\institution{Pisa}{Scuola Normale Superiore, Piazza dei Cavalieri 7, I-56126 Pisa, Italy}

\title{
Gauge gravity duality for $d$-wave
superconductors:
prospects and challenges
}

\authors{
Francesco~Benini\worksat{\PU,}\footnote{e-mail: {\tt fbenini@Princeton.EDU}},
Christopher~P.~Herzog\worksat{\PU,}\footnote{e-mail: {\tt cpherzog@Princeton.EDU}},
Rakibur~Rahman\worksat{\Pisa,}\footnote{e-mail: {\tt rakibur.rahman@sns.it}},
Amos~Yarom\worksat{\PU,}\footnote{e-mail: {\tt ayarom@princeton.edu}}
}

\abstract{
We write down an action for a charged, massive spin two field in a fixed Einstein background.
Despite some technical problems,
we argue that in an effective field theory framework and in the context of the AdS/CFT correspondence, this action
can be used to study the properties of a superfluid phase transition with a $d$-wave order parameter in a dual strongly interacting field theory. We investigate the phase diagram and the charge conductivity of the superfluid phase.
We also explain how possible couplings between the spin two field and bulk fermions affect the fermion spectral function.
}

\date{June 2010}

\maketitle

\tableofcontents

\section{Introduction and summary}

In the context of gauge gravity duality \cite{Maldacena:1997re,Gubser:1998bc,Witten:1998qj}, a holographic superconductor has come to mean a strongly interacting field theory that undergoes a superconducting or superfluid phase transition and that has in addition a dual gravitational description.  See \cite{Hartnoll:2009sz,Herzog:2009xv,Horowitz:2010gk} for reviews.
Holographic superconductors are interesting because they describe
condensed phases of strongly coupled, planar, gauge theories. Also,
they may give some insight into real world strongly interacting
superconductors and superfluids such as the cuprates and helium-3.
To date, most of the holographic examples have had either $s$-wave or $p$-wave order parameters  (three exceptions are \cite{Chen:2010mk,Herzog:2010vz,Zeng:2010fx}).
Because of
a possible relationship with the cuprates,
in this paper we explore a holographic toy-model for the condensation of a $d$-wave order parameter.

A toy-model for an $s$-wave holographic superconductor was first constructed in \cite{Hartnoll:2008vx} following the observation in \cite{Gubser:2008px} that in AdS space black holes can have scalar hair. A more realistic construction of an $s$-wave holographic superconductor dual to an orbifold of $\mathcal{N}=4$ SYM was discussed in \cite{Gubser:2009qm}. See \cite{Gauntlett:2009dn} for a similar construction using an $M$-theory truncation and
\cite{Ammon:2008fc}
for a $p$-wave construction using probe D-branes.

Comparing a holographic superconductor to a high $T_c$ superconductor or other materials is tempting but difficult:  For a gauge theory such as ${\mathcal N}=4$ $SU(N)$ Super Yang-Mills, the dual gravitational description is most useful in its strongly coupled and large $N$ limit. Moving away from this limit is challenging. The role of supersymmetry and the absence of a lattice are two other important issues that must be confronted before any detailed comparison can be made.
Nevertheless, the need for non-perturbative insight in studying high $T_c$ materials
\cite{Carlson:2002}
suggests an exploration of holographic methods.

In the holographic setup, a $d$-wave order parameter is dual to a charged massive spin two field propagating in an asymptotically AdS geometry.
While the action for a neutral, massive spin two field in a flat background has been known for
over seventy years \cite{Fierz:1939ix}, finding an action for a charged spin-two field in a curved spacetime is an unsolved problem.
In section \ref{S:Spin2} we discuss the perils and pitfalls in constructing such an action.
We settle for a relatively simple action with some known problems that we believe can be fixed through the addition of higher dimensional operators.  Thus, we work in an effective field theory framework.

With 
the action for a charged massive spin two field in AdS space in hand, we can solve the bulk equations of motion and study the resulting condensed phase in the boundary theory. This is carried out in section \ref{S:solution}.
For a particular ansatz in which only a single, real component of the spin two field is turned on, and after an appropriate field redefinition, we observe that the equations of motion for the spin two field are identical to those of the charged massive scalar field studied in \cite{Hartnoll:2008vx}. Thus, we are guaranteed that when turning on a chemical potential there will exist a critical temperature below which the spin two field condenses.

We proceed to compute the conductivity of the $d$-wave superconductor in section \ref{S:conductivity} and to study fermion correlators in this background in section \ref{S:fermions}. As was the case for the massive charged scalar, we find that the conductivity associated with the superconductor develops a gap-like feature (depicted in figure \ref{F:conductivity})
but it also has a sharp $\delta$-function spike indicative of a bound state.
In studying the spectral function for fermions, we find that for a particular coupling between the fermions and the spin two field, the spectral function has a band-like structure in addition to the $d$-wave gap and Fermi-arcs observed in \cite{Benini:2010qc}.\footnote{%
See \cite{Vegh:2010fc} for a related discussion of Fermi arcs in $p$-wave superconductors.
}
 
\section{Spin two fields}
\label{S:Spin2}

One way of generating a $d$-wave condensate in the boundary theory is to have a massive, charged spin two field condense in an asymptotically AdS${}_{d+1}$ geometry. By a massive spin two field in $d+1$ spacetime dimensions, we mean a field that transforms locally in the $(d+2)(d-1)/2$ dimensional irreducible representation of the little group $SO(d)$ of the Lorentz group $SO(1,d)$.  If such a field is represented by a symmetric tensor $\varphi_{\mu\nu}$ then there should be enough constraint equations to eliminate $d+2$ of its $(d+1)(d+2)/2$ components. In flat space, the Fierz-Pauli Lagrangian \cite{Fierz:1939ix} does this job. It  is given by
\begin{equation}
\label{E:FP}
\mathcal{L} = \frac{1}{4} \Big[
- \partial_\rho \varphi_{\mu\nu} \partial^\rho \varphi^{\mu\nu} + 2 \varphi_\mu \varphi^\mu - 2 \varphi^\mu \partial_\mu \varphi + \partial_\mu \varphi \partial^\mu \varphi - m^2 ( \varphi_{\mu\nu} \varphi^{\mu\nu} - \varphi^2) \Big] \;,
\end{equation}
where we have introduced the notation $\varphi_\rho \equiv \partial^\mu \varphi_{\mu\rho}$ and $\varphi \equiv \varphi^\mu_\mu$. The equations of motion (EOM's) and
constraints following from \eqref{E:FP} are
\begin{subequations}
\label{E:FPequations}
\begin{align}
0 &= \left(\square - m^2 \right) \varphi_{\mu\nu} \ , \\
\begin{split}
\label{E:FPconstraints}
0 &= \partial_{\mu} \varphi^{\mu\nu} \ , \\
0 &= \varphi \;,
\end{split}
\end{align}
\end{subequations}
which indeed give the correct number of degrees of freedom. We reproduce \eqref{E:FPequations} in appendix \ref{A:FPEOM}.

Naively, one might think that covariantizing \eqref{E:FP} and introducing a minimal coupling to a $U(1)$ field will yield a consistent action for a massive charged spin two field in a curved background. Unfortunately such a Lagrangian does not produce appropriate constraint
equations as in \eqref{E:FPconstraints} and the spurious propagating degrees of freedom  give rise to pathologies such
as ghosts, loss of hyperbolicity and faster than light travel.
There have been several attempts to deal with these issues in the literature.  We mention two. The Buchbinder-Gitman-Pershin (BGP) Lagrangian \cite{Buchbinder:2000fy} describes a  neutral massive spin two field in an Einstein manifold, and has the appropriate number of degrees of freedom that propagate causally. The Federbush Lagrangian constructed in \cite{Federbush:1961} describes a charged massive spin two particle
propagating in flat space. It has the correct number of propagating degrees of freedom, but also has the unwanted feature of generating superluminal modes \cite{Velo:1972rt}.

The Lagrangian density we will use has the following form:
\begin{equation}
\begin{split}
\label{E:ActionSimp}
\mathcal{L} &= - |D_\rho \varphi_{\mu\nu}|^2 + 2|D_\mu \varphi^{\mu\nu}|^2 + |D_\mu \varphi|^2 - \big[ D_\mu \varphi^{*\mu\nu} D_\nu \varphi + \text{c.c.} \big] - m^2 \big( |\varphi_{\mu\nu}|^2 - |\varphi|^2 \big) \\
&\quad +2  R_{\mu\nu\rho\lambda} \varphi^{*\mu\rho} \varphi^{\nu\lambda}
- R_{\mu\nu} \varphi^{*\mu\lambda} \varphi^\nu_\lambda
- \frac{1}{d+1} R | \varphi |^2
- i  q F_{\mu\nu} \varphi^{*\mu\lambda} \varphi^\nu_\lambda - \frac14 F_{\mu\nu} F^{\mu\nu} \;,
\end{split}
\end{equation}
where we have introduced $D_\mu = \nabla_\mu - i q A_\mu$ when acting on $\varphi_{\mu\nu}$ and now $\varphi_\rho = D^\mu \varphi_{\mu\rho}$.
The EOM's which follow from (\ref{E:ActionSimp}) are
\bea
\label{EOM probe action}
& 0 = (\square - m^2) \varphi_{\mu\nu} - 2 D_{(\mu} \varphi_{\nu)} + D_{(\mu} D_{\nu)} \varphi - g_{\mu\nu} \big[ (\square - m^2) \varphi - D^\rho \varphi_\rho \big] \\
&\qquad + 2 R_{\mu\rho\nu\lambda} \varphi^{\rho\lambda} - g_{\mu\nu} \frac {R}{d+1} \varphi - i\frac q2 \big( F_{\mu\rho} \varphi^\rho_\nu + F_{\nu\rho} \varphi^\rho_\mu \big) \\
& D_\mu F^{\mu\nu} = J^\nu
\eea
where
\be
J^\nu = i \varphi^*_{\alpha\beta} (D^\nu \varphi^{\alpha\beta} - D^\alpha \varphi^{\nu\beta}) + i(\varphi^*_\alpha - D_\alpha \varphi^*)(\varphi^{\nu\alpha} - g^{\nu\alpha} \varphi) + \text{h.c.} \;.
\ee

The Lagrangian \eqref{E:ActionSimp} has the following desirable properties:
\begin{itemize}
\item
It has familiar limits. For a neutral, non-interacting spin two field in flat spacetime, it reduces to the Fierz-Pauli Lagrangian \cite{Fierz:1939ix}.  For a neutral spin two field in a fixed Einstein background, ${\mathcal L}$ reduces to that of BGP \cite{Buchbinder:2000fy}.
For a charged spin two field in flat spacetime, ${\mathcal L}$ reduces to the Lagrangian of Federbush \cite{Federbush:1961}.

\item
It is ghost-free, and for generic values of $F_{\mu\nu}$, it describes $(d+2)(d-1)/2$ propagating degrees of freedom.  We demonstrate this fact in Appendix \ref{app:constraints}.

\item
The Lagrangian is unique in an appropriate sense:
Consider a spin two Lagrangian with all possible operators up to dimension $d+1$ and quadratic in $\varphi_{\mu\nu}$. If we choose the couplings of this Lagrangian to be different from those in \eqref{E:ActionSimp}, then some of the would-be constraint equations become dynamical, introducing ghosts. We provide more details about the uniqueness properties of \eqref{E:ActionSimp} in appendix \ref{app:constraints}.
\end{itemize}

At the same time, our Lagrangian has a number of limitations:
\begin{itemize}
\item
As was the case for the BGP Lagrangian, we are restricted to work in a fixed background spacetime that satisfies the Einstein condition
\begin{equation}
\label{EinsteinCond}
R_{\mu\nu} = \frac{2 \Lambda}{d-1} \, g_{\mu\nu}  \;.
\end{equation}
Otherwise, the constraint equations will not be satisfied and we will have too many propagating degrees of freedom.
In the context of holographic superconductors,
this restriction forces us to work in the probe limit 
\cite{Hartnoll:2008vx}
where the spin two field and gauge field do not backreact on the metric; since the metric will not be perturbed by the matter content of the theory, it will automatically satisfy \eqref{EinsteinCond}.

\item
Similar to the EOM's following from the Federbush Lagrangian \cite{Federbush:1961}, the EOM's following from \eqref{E:ActionSimp}, for generic values of $F_{\mu\nu}$, are either non-hyperbolic or lead to non-causal propagation \cite{Velo:1972rt}.  Fortunately, these effects are small in the sense that they disappear in the limit where the quantities
\begin{equation}
\label{cutoff}
\frac{q |F_{\ulambda\unu}|}{m^2} \;, \qquad \frac{q  |F_{\ulambda\unu;\, \usigma}|}{m^3}
\end{equation}
vanish. (Underlined indices are in a local vielbein basis.)
As we summarize in section \ref{S:validity}, there are a number of reasons to believe that one can correct these problems by adding to the Lagrangian terms that are higher order in $F_{\mu\nu}$.%
\footnote{One of the reasons which leads us to believe that these corrections exist is the existence of the Argyres-Nappi Lagrangian \cite{Argyres:1989cu}. The Argyres-Nappi Lagrangian is a consistent causal Lagrangian for a massive charged spin two particle in a constant electromagnetic background and 26 flat spacetime dimensions.
Further indication that higher order terms containing $F_{\mu\nu}$ are needed comes from the fact that the Lagrangian (\ref{E:ActionSimp}), when rewritten in terms of St\"uckelberg fields and in a gauge that makes the kinetic terms canonical, contains such terms \cite{Porrati:2008an,Porrati:2008ha}. We will come back to this point in section \ref{S:validity}.}
Such a series expansion is sensible only in a frame where these violations of causality are small. In section \ref{S:validity} we will discuss in what regime of parameter-space these corrections can be made small.

\end{itemize}

\section{The bulk dual of a $d$-wave condensate}
\label{S:solution}

In this section we adapt the standard AdS/CFT map to a massive spin two field, and study the gravitational solution corresponding to a holographic $d$-wave superconductor.

\subsection{The conformal dimension of the boundary operator.}
\label{S:AdSCFTspin2}

Since AdS space is an Einstein manifold, we can use the Lagrangian $\mathcal{L}$ defined in \eqref{E:ActionSimp} to describe a charged spin two field propagating in AdS space. The gauge gravity duality implies that this Lagrangian also describes a spin two operator $\cO_{mn}$ on the boundary theory. Since $\mathcal{L}$ was not derived from a higher dimensional string action, it is not clear what the boundary theory dual to \eqref{E:ActionSimp} is, or if it exists as a proper field theory. In this sense the action \eqref{E:ActionSimp} should be thought of as a toy-model. Assuming the boundary theory exists, we can use the AdS/CFT dictionary described in \cite{Gubser:1998bc,Witten:1998qj} to make some general observations regarding the dual boundary theory spin two operator. The following analysis of the correspondence for a spin two field generalizes the well known duality between the energy momentum tensor of the boundary theory and the graviton---a neutral, massless spin two field in the bulk.

Consider the spin two Lagrangian \eqref{E:ActionSimp} for a neutral spin two field propagating in an AdS background with line element
\begin{equation}
\label{standard AdS}
ds^2 = \frac{L^2}{z^2} \, \big( -dt^2 + d\vec x^2 + dz^2 \big) \;.
\end{equation}
As written, it is not clear from \eqref{E:ActionSimp} how to define the mass of the spin two field, in the sense that one can absorb couplings between the spin two field and the curvature into a redefinition of $m^2$ \cite{Deser:1983mm}. We resolve this ambiguity by defining the mass squared of the spin two field, $m^2$, such that when $m^2=0$ the action has an enhanced gauge symmetry
of the form $\varphi_{\mu\nu} \to \varphi_{\mu\nu} + \delta \varphi_{\mu\nu}$ with $\delta\varphi_{\mu\nu} = D_\mu \xi_\nu + D_\nu \xi_\mu$, where $\xi_\mu$ is an infinitesimal parameter.
This gauge invariance is unique to the graviton, and is associated with diffeomorphisms.

In what follows we will be setting $m^2 \neq 0$. When $m^2 \neq 0$ and $m^2 \neq \frac{d-1}{d(d+1)}R$,
the equations of motion and constraints derived from $\mathcal{L}$ are similar to the flat spacetime case (\ref{E:FPequations}):
\begin{subequations}
\begin{align}
0 &= \big( \square - m^2 \big) \varphi_{\mu\nu} + 2 R_{\mu\lambda\nu\rho} \varphi^{\lambda\rho} \\
\begin{split}
\label{E:Bconstraints}
0 &= D^\mu \varphi_{\mu\nu} \\
0 &= \varphi^\mu_\mu \;.
\end{split}
\end{align}
\end{subequations}
Curiously, when $m^2 = \frac{d-1}{d(d+1)}R$ the spin two action has an enhanced gauge symmetry of the form $\delta \varphi_{\mu\nu} = D_\mu D_\nu \xi + g_{\mu\nu} \frac R{d(d+1)} \xi$. This is related to the existence of a partially massless spin-2 field \cite{Deser:2001pe}.

For operators with spin $s > 0$, bulk fields and boundary operators have a different number of components. In what follows, we will argue that the 
equations of motion for a bulk spin two field $\varphi_{\mu\nu}$
have two linearly independent solutions, and that the number of integration constants is twice the number of components of a spin two operator $\cO_{mn}$ in the boundary theory. In \ref{S:dwaveCondensate} we will identify these solutions with the source and expectation value of the spin two operator.

The explicit form of the constraint equations (\ref{E:Bconstraints}) is
\bea
\label{E:constraintsAdS}
\big[ (d-1) - z \partial_z \big] \varphi_{mz} &= z \, \eta^{kn} \partial_{k} \varphi_{nm} \\
\big[ (d-1) - z \partial_z \big] \varphi_{zz} &= z \, \eta^{kn} \partial_{k} \varphi_{nz} \\
\eta^{kn}\varphi_{kn} + \varphi_{zz} &= 0 \;,
\eea
where roman indices run over the boundary directions $0,\ldots,d-1$ and $\eta^{kn}$ is the Minkowski metric. We observe that the components of $\varphi_{\mu\nu}$ with one or more legs along the radial direction are completely determined by the components of $\varphi_{\mu\nu}$ with both legs along the boundary directions. Furthermore, the near boundary series expansion of $\varphi_{\mu z}$ starts off with a higher power of $z$ than the near boundary series expansion of $\varphi_{mn}$.
Therefore, the last equation in \eqref{E:constraintsAdS} implies that
not only is $\varphi_{\mu\nu}$ traceless, 
but that the leading term in a near boundary series expansion of $\varphi_{mn}$ is also
traceless. Given the relation between $\langle {\mathcal O}_{mn} \rangle$ and $\varphi_{mn}$ which we make precise below, tracelessness  of the leading term of a near boundary series expansion of $\varphi_{mn}$ ensures that the boundary operator 
$\cO_{mn}$ is also traceless.\footnote{%
The equation of motion for the spin two field is a second order differential equation. Tracelessness of the leading term in a series expansion of the solution applies to both
of the two linearly independent solutions. Thus, both the source term and the expectation value of the dual operator will be traceless.}

The equations of motion for the remaining components of the spin two field are
\be
\label{E:spin2mneom}
0 = - \big[ 2(d-2) + L^2 m^2 \big] \varphi_{mn} - 2 \eta_{mn} \eta^{kp}\varphi_{kp} + z^{d-3} \partial_z \big( z^{5-d} \partial_z \varphi_{mn} \big) + z^2 \partial^2 \varphi_{mn} - 4z \partial_{(m} \varphi_{n)z} \;.
\ee
These equations have two solutions whose near boundary behavior is given by $\varphi_{mn} \sim z^{\Delta - 2}$ and $\varphi_{mn} \sim z^{d-\Delta-2}$, where $\Delta$ is defined through
\be
\label{mass dimension formula}
m^2 L^2 = \Delta(\Delta - d) \;.
\ee
The parameter $\Delta$ should be identified with the dimension of the operator $\cO_{mn}$ as we now explain. Using the AdS/CFT dictionary, the coefficient of the $z^{\Delta -2}$ term can be associated with an expectation value, $\langle \cO_{mn} \rangle$, while the coefficient of the $z^{d-\Delta-2}$ term serves as a source, $\cO^{(s)}_{mn}$. Since $\varphi_{\mu\nu}$ has definite scaling behavior under $z \to z/\lambda$, the ratio $\langle \cO_{mn} \rangle / \cO^{(s)}_{mn}$ must have scaling dimension $2 \Delta - d$.  This ratio, through the theory of linear response, is the Fourier transform of a retarded Green's function which, for an operator of dimension $\Delta$, will also have scaling dimension $2 \Delta -d$.\footnote{%
As a rule of thumb, the two near-boundary behaviors of the boundary components $\Psi_{m_1 \dots m_p}$ of a generic tensor field $\Psi_{\mu_1 \dots \mu_p}$ with covariant action and in the standard coordinates (\ref{standard AdS}), are $z^{\Delta - p}$ and $z^{d-\Delta - p}$, where $\Delta$ is the dimension of the boundary operator $\cO_{m_1\dots m_p}$. This scaling can be motivated by noticing that the boundary and bulk metrics are related by the rescaling $g^\text{bnd} = z^2 g^\text{bulk}\big|_{\text{bnd}}$. Thus, the identification of tensor quantities expressed in a boundary or bulk frame requires them to be rescaled accordingly.}
As expected, when $\Delta=d$ (which is the protected dimension of the boundary theory stress tensor) then $m^2=0$.

The dimension of a spin two primary operator in $d$ spacetime dimensions is constrained by unitarity and conformal symmetry. An analysis of the positive-energy unitary representations of the three dimensional conformal group $SO(3,2)$ was first carried out in \cite{Evans:1967}. The analysis has been generalized to the four dimensional conformal group $SO(4,2)$ in \cite{Mack:1975je} and to $d$ dimensions in \cite{Minwalla:1997ka}.
In short, the analysis in \cite{Evans:1967,Mack:1975je,Minwalla:1997ka} states that the positive-energy representations of $SO(d,2)$ are classified by their quantum numbers in $SO(2)\times SO(d)$, which are the dimension $\Delta$ of the primary operator in the module and its spin. Unitarity sets a spin-dependent lower bound on $\Delta$.
In particular, a totally symmetric and traceless tensor operator $\cO_{m_1 \dots m_s}$ of spin $s \geq 1$ must satisfy $\Delta \geq d -2 +s$. When this bound is saturated, then $\cO_{m_1 \dots m_s}$ is a conserved current in the sense that $\partial^{m_1} \cO_{m_1 \dots m_s}=0$. For spin two operators the unitarity bound implies that $\Delta \geq d$. Thus, our bulk spin two fields should have a positive mass squared, $m^2 \geq 0$. In \cite{Deser:2001pe, Deser:2001us, Deser:2001wx} (see also \cite{Dolan:2001ih}) it was shown that also from the bulk point of view, stability implies that $m^2\geq 0$. In
appendix \ref{app:BF bound} we discuss how to obtain such a bound on the mass by generalizing the analysis of Breitenlohner and Freedman \cite{Breitenlohner:1982bm}.

\subsection{Generating a $d$-wave condensate}
\label{S:dwaveCondensate}

The spin two action we consider describes the correct number of degrees of freedom for a massive spin two field only on an Einstein background. Therefore, we need a control parameter such that the bulk stress tensor can be made negligible when compared to the cosmological constant in the Einstein equations. This regime of the theory, called the probe limit in \cite{Hartnoll:2008vx}, is achieved by taking the charge of the spin two field $q$ to be large. Indeed, rescaling the spin two field and the gauge field by a factor of $q$, $\varphi_{\mu\nu}=\tilde\varphi_{\mu\nu} /q $ and $A_{\mu} = \tilde A_\mu / q $, the Lagrangian (\ref{E:ActionSimp}) gets rescaled by a factor of $q^2$, $\cL = \tilde \cL/q^2$ (where $\tilde \cL$ is independent of $q$). In the limit in which $q$ is infinite but $\tilde{A}_{\mu}$ and $\tilde{\varphi}_{\mu\nu}$ are kept fixed, the bulk stress tensor goes to zero. We will always work in this limit, expressing boundary quantities in units of $q\mu$, where $\mu$
is the chemical potential in the boundary theory.

To look for a thermal state on the boundary theory in which the spin two field presumably condenses, we need to solve the equations of motion \eqref{EOM probe action} in a black hole background which satisfies the Einstein equation \eqref{EinsteinCond}. One such geometry is given by the black brane solution
\begin{equation}
ds^2 = \frac{L^2}{z^2} \Big(-f(z)\,dt^2 + d \vec x_{d-1}^2 + \frac{dz^2}{f(z)} \Big)
\end{equation}
where
\begin{equation}
f(z) = 1 - \Big( \frac{z}{z_h} \Big)^d \;.
\end{equation}
The black hole horizon is located at $z=z_h$, while the conformal boundary of the spacetime is located
at $z=0$.
The Schwarzschild temperature of this black hole is
\begin{equation}
T = \frac{d}{4 \pi z_h} \;.
\end{equation}

We consider an ansatz where $\varphi_{\mu\nu}$ and $A_\mu$ depend only on the radial coordinate $z$ and where only the space components of $\varphi_{\mu\nu}$ are turned on.
For $d=3$,
since $\varphi_{\mu\nu}$ is a complex field, we can use a spatial $(xy)$ rotation combined with a $U(1)$ gauge transformation to choose one of the two components of $\varphi$, $\varphi_{xx} = -\varphi_{yy}$ or $\varphi_{xy}$, to be real and the other one to be imaginary.
However, similar to what was found in \cite{Hartnoll:2008vx}, it is consistent to turn on a single real component of $\varphi$ and also to set all the components of the gauge field except for $A_t$ to be zero.  Our ansatz is then
\begin{equation}
\label{ansatz}
A = A_\mu \, dx^\mu \equiv  \phi(z) \, dt  \;, \qquad\qquad
\varphi_{xy}(z) \equiv \frac{L^2}{2z^2} \, \psi(z) \;,
\end{equation}
with all other components of $\varphi_{\mu\nu}$ set to zero, and $\phi$ and $\psi$ real. The more general ansatz in $d=3$ with one mode real and the other one imaginary could lead to interesting physics and we leave it for future work. The ansatz \eqref{ansatz} satisfies $\varphi = \varphi_\mu = F_{\mu\rho} \varphi^\rho_\nu = 0$. Instead of turning on $\varphi_{xy}$ in \eqref{ansatz} we could have considered a non-vanishing value for $\varphi_{xx-yy} \equiv \varphi_{xx} = - \varphi_{yy}$. These two ans\"{a}tze are equivalent under a
$\pi/4$ rotation. Indeed, under a counterclockwise rotation in the $xy$-plane by an angle $\theta$, the vector $\big( \begin{smallmatrix} \varphi_{xx-yy} \\ \varphi_{xy} \end{smallmatrix} \big)$ transforms under the action of the matrix
\be
\label{condensate rotation}
R_\theta = \mat{ \cos 2\theta & -\sin 2\theta \\ \sin 2\theta & \cos 2\theta } \;.
\ee
Note that our ansatz continues to satisfy the equations of motion when $d>3$, but the most general ansatz will require turning on more components of $\varphi_{\mu\nu}$.

Using the ansatz (\ref{ansatz}), the EOM's for $\phi$ and $\psi$ are
\begin{align}
\label{E:EOMphi}
0 &= \phi'' + \frac{3-d}z \, \phi' - \frac{q^2L^2}{z^2 f} \, \psi^2 \, \phi \\
\label{E:EOMpsi}
0 &= \psi'' + \left( \frac{f'}{f} - \frac{d-1}{z} \right) \psi' + \left( \frac{ q^2 \phi^2}{f^2} - \frac{m^2 L^2}{z^2 f} \right)  \psi \;,
\end{align}
where a prime denotes a derivative with respect to the radial coordinate $z$. Near the asymptotically AdS boundary the series expansion for $\psi$ takes the form:%
\footnote{Up to possible logarithms for special values of $\Delta$.}
\begin{equation}
\label{E:psiexpansion}
\psi(z) =  z^{d-\Delta} \big[ \psi^{(s)}  + O(z) \big] + z^{\Delta} \Big[ \frac{\langle \cO_{xy} \rangle}{2 \Delta - d}   + O(z) \Big] \;,
\end{equation}
where $\Delta$ is the conformal dimension of the boundary spin two operator defined in (\ref{mass dimension formula}). The $O(z)$ terms represent subleading corrections in powers of $z$ which can be computed perturbatively from the EOM's. The values $\psi^{(s)}$ and $\langle \cO \rangle$ are integration constants which need to be determined.

The boundary conditions we impose on the spin two field are that the modulus $\varphi_{\mu\nu} \varphi^{\mu\nu}$ is finite at the black hole horizon, which implies that $\psi$ is finite. We also require that $ \psi^{(s)}=0$ so that the spin two field is not sourced. The gauge field $A_t$ is required to vanish at the horizon. We interpret the value of $A_t$ at the asymptotically AdS boundary as the chemical potential $\mu$. The equations \eqref{E:EOMphi} and \eqref{E:EOMpsi} are identical to the EOM's for the Abelian Higgs model in AdS${}_{d+1}$ \cite{Hartnoll:2008vx}. Thus, a condensed solution exists below a certain critical temperature $T_c$: besides the ``normal phase'' solution
\begin{equation}
\label{E:normalphase}
A = \phi(z) \, dt = \mu \, \Big[ 1 - \Big( \frac{z}{z_h} \Big)^{d-2} \Big] \, dt  \;, \qquad\qquad
\psi = 0
\end{equation}
which exists for all values of the chemical potential, an additional solution with $\psi \neq 0$ exists whenever $T/q\mu$ is small enough. If we consider the chemical potential as fixed, then the critical value of $T/q\mu$ defines a critical temperature $T_c$ below which the spin two field condenses.

To determine $T_c$, we look for the temperature at which the normal phase solution \eqref{E:normalphase} becomes unstable to perturbations in the sense that $\psi$ develops a normal mode.  There exist several strategies for determining $T_c$.
For large $m^2$ the problem becomes classical, and the critical temperature can be approximated by determining the condition for a global minimum away from the horizon
in the potential
of a charged, massive particle in the normal phase.  We examine this method in Appendix \ref{app:semiclassics}.
To determine $T_c$ for other values of the mass, and to find an explicit solution for $T< T_c$, we must resort to numerical techniques.

\begin{figure}
\begin{center}
\includegraphics[height=8cm]{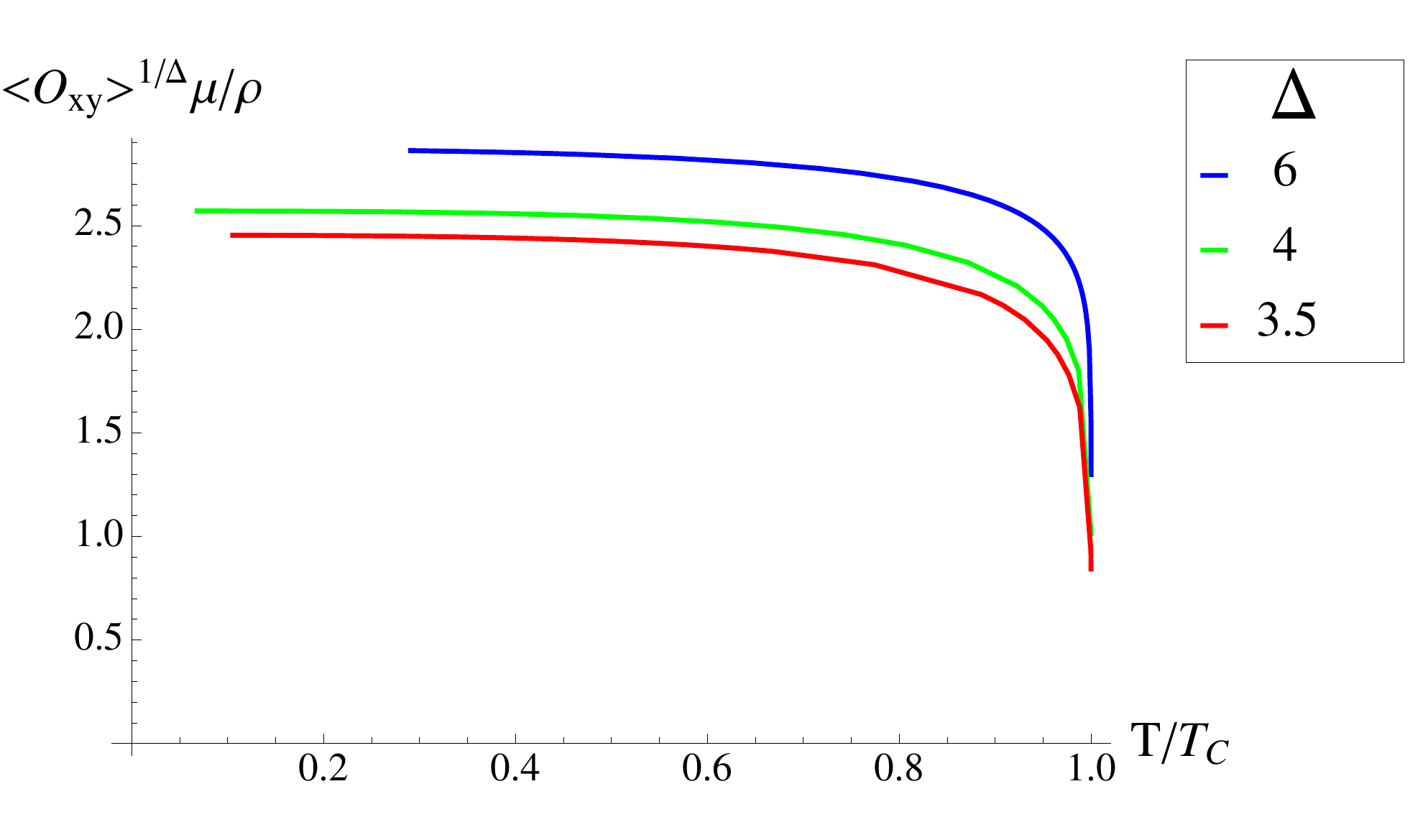}
\caption{(Color online) The order parameter $\langle \cO_{xy} \rangle$ in $d=3$ spacetime dimensions as a function of the temperature for various values of $\Delta$ which decrease from top to bottom. The chemical potential $\mu$ and the charge density $\rho$ were obtained numerically using standard methods. The units we use to measure $\langle \cO_{xy} \rangle$ are similar to those used in \cite{Yarom:2009uq}.
\label{F:typicalO}}
\end{center}
\end{figure}

With a numerical solution to (\ref{E:EOMphi}) and (\ref{E:EOMpsi}) in hand, one may compute the expectation value of the boundary operator $\cO_{mn}$ dual to the spin two field $\varphi_{mn}$ via the standard AdS/CFT rules with flat Cartesian coordinates on the boundary \cite{Gubser:1998bc,Witten:1998qj}:
\begin{equation}
\langle {\cO}_{mn}  \rangle = \lim_{z \to 0} \frac{\delta S}{\delta  \varphi_{mn}^{(s)}}\,,
\end{equation}
where $\varphi_{mn}^{(s)} = \lim_{z\to 0} z^{\Delta-d+2} \varphi_{mn}(z)$ is the source term, and $S$ is the on-shell action plus possible boundary terms.%
\footnote{In \cite{Klebanov:1999tb} a different normalization for $\psi^{(s)}$ was suggested. This will result in a rescaling of $\langle \cO_{xy} \rangle$ by an overall $\Delta$-dependent multiplicative factor.}
Instead of working out the boundary counter terms for the spin two field and computing $\langle \cO_{mn} \rangle$ directly, we will use the similarity between the spin two system and the Abelian Higgs model in AdS$_{d+1}$ which was mentioned earlier.
It is straightforward to show that the on-shell Abelian Higgs action discussed in \cite{Hartnoll:2008vx}
is identical to the one following from \eqref{E:ActionSimp} and the ansatz \eqref{ansatz}.
The only counter term which is required to render the on-shell Abelian Higgs action finite is a mass counter term. One can show that such a counter term is equivalent to adding a boundary counter term proportional to $|\varphi_{\mu\nu}|^2$ to \eqref{E:ActionSimp}.
A typical dependence of $\langle \cO_{xy}\rangle$ on the temperature, obtained numerically, is shown in figure \ref{F:typicalO}.

We conclude by stressing that in the probe limit it is not possible to reach zero temperature when $m^2>0$. In the probe limit, the background generated by the spin two field is the same as that of the Abelian Higgs model. The zero temperature limit of the Abelian Higgs model with backreaction was analyzed in \cite{Gubser:2009cg,Horowitz:2009ij} where it was shown that when $m^2>0$ the backreacted geometry has Lifshitz symmetry. In the probe limit, in contrast, the geometry has conformal symmetry when the temperature is zero. Thus, at zero temperature the backreaction of the fields produces a qualitative difference on the geometry.
Still, for any positive temperature $T$, it is always possible to take $q$ large enough such that the stress tensor is negligible.%
\footnote{
An additional problem with the $T\to 0$ limit at non-zero density is that, although the action near $T_c$ has some claim to universality, at lower temperatures irrelevant operators become important and then the simple action we have written down needs to be corrected.}

\subsection{Regime of validity of the effective Lagrangian}
\label{S:validity}

In equation \eqref{cutoff} we claimed that the field strength and its derivatives
must be small when measured in units of the mass of the spin two field
in order for our spin two action \eqref{E:ActionSimp} to be valid as an effective theory. More precisely, we require
all the components of $F_{\lambda\nu}$ in a local vielbein basis (which we denote $F_{\ulambda\unu}$)
to be small.%
\footnote{This statement is frame-dependent. The existence of a frame where all field strength components are small assures that the difference between our effective action and the full causal action (that we assume to exist) organizes in a series expansion.}
In the AdS-Schwarzschild background where $A_t$ is given by (\ref{E:normalphase}) the only non-vanishing component of the field strength is
\begin{equation}
\label{cutoffapplied}
\frac{q F_{\ut\uz}
}{m^2} = \frac{(d-2) q\mu z^{d-1}}{\Delta (\Delta - d) z_h^{d-2}} \;.
\end{equation}
At any non-zero temperature the coordinate $z$ can be no larger than $z_h$, yielding the condition
\begin{equation}
\label{E:RNtrust}
\frac{(d-2) q\mu z_h}{ \Delta (\Delta - d)} \ll 1 \;.
\end{equation}
For large $m^2$ and $d=3$ appendix \ref{app:semiclassics} provides an estimate of the critical temperature below which the spin two field condenses.  In order for the phase transition to occur in a regime where we can still trust the spin two action we find that \eqref{E:RNtrust} and \eqref{E:TcAnalytic} imply
\begin{equation}
\label{E:Suppression}
\Delta \gg 1 \ .
\end{equation}
By the same analysis, terms involving the gradient of $F_{\mu\nu}$ in (\ref{cutoff}) will be suppressed by an additional power of $\Delta$.

In principle effects of non-hyperbolicity and non-causality can be made small enough by working with a spin two field of very large conformal dimension. As we now explain, in practice it is difficult to access the solutions numerically when $\Delta$ is large. The requirement that the boundary spin two operator is not sourced implies that the leading $z^{d-\Delta-2}$ term in the series expansion of the bulk spin two field must vanish. The expectation value of the condensate is proportional to the $z^{\Delta-2}$  term in the series expansion of the bulk spin two field. Thus, when shooting from the horizon to the boundary, one needs to set the source term to zero to very high precision in order for the value of the expectation value to be accurate.  In our numerical work, we chose a moderate value for $\Delta$ where we feel confident that the numerical error is small. We hope that the results will not qualitatively change with a larger $\Delta$ or with corrections to the action.

Unfortunately,
our expansion in inverse powers of $\Delta$
implies that dimension six operators of the schematic form
\begin{equation}
\frac{q F}{m^2} \, D \varphi^* D\varphi
\label{dim6ops}
\end{equation}
will not be suppressed relative to the dimension four operator
$F \varphi^* \varphi$ that we did consider in constructing \eqref{E:ActionSimp}.
One way to motivate the addition of such dimension six operators is that they appear in the Argyres-Nappi action \cite{Argyres:1989cu}. Using bosonic string theory, Argyres and Nappi found a consistent action for a spin two field in flat spacetime and constant electromagnetic background.
If one expands the Argyres-Nappi action to linear order in $F_{\mu\nu}$, one finds terms precisely of the form (\ref{dim6ops}).%
\footnote{The action is only consistent in $d=26$ dimensions. For $d\neq 26$,
a constraint equation becomes
dynamical through the appearance of terms quadratic
in $F_{\mu\nu}$.  Thus, the action has the correct number of degrees of freedom and is causal only up to $O(F^2)$. It is not presently known whether a modification of the original action can work in arbitrary dimension.}

The terms in \eqref{dim6ops} have a
benefit.  Using the effective Lagrangian \eqref{E:ActionSimp}, the gyromagnetic coupling $g$ of the spin two field takes a surprising value of 1/2.  Physical arguments suggest that $g=2$ \cite{Ferrara:1992yc, Deser:2001dt}. The reason for this mismatch is that the gyromagnetic coupling is affected by irrelevant corrections and cannot be read off from our leading order Lagrangian; corrections of the form \eqref{dim6ops} alter $g$ and the value $g=2$ can be easily obtained.

So why don't we include these dimension six operators? The honest answer is that we currently do not have a prescription for fixing their coefficients uniquely and that they make the calculations much more complicated.
However, we can show that, for the most part, our results will be independent of these terms.
With the ansatz \eqref{ansatz}, we have the relations $F_{\mu\nu} \varphi^\nu_\rho = 0$ and $\varphi_{\mu}=0$. These equalities imply that the problematic dimension six operators in \eqref{dim6ops} won't affect the condensed solution described in section \ref{S:dwaveCondensate}. Similarly, these operators won't affect computations of fermion correlators in the condensed phase performed in \cite{Benini:2010qc} and discussed in more detail in section \ref{S:fermions}. However, for the conductivity calculation below, these operators will probably play a role. Given the added complexity of the equations one must solve, we content ourselves by studying the simpler action (\ref{E:ActionSimp}) and leave a detailed investigation of the operators (\ref{dim6ops}) for the future.

\section{Conductivity}
\label{S:conductivity}

The optical conductivity $\sigma(\omega)$ associated with the spin two condensate is a measure of the response of the (weakly gauged) $U(1)$ current to a time varying external source,
\begin{equation}
\label{E:sigmadef}
\vec{J} = \sigma \cdot \vec{E} \;,
\end{equation}
where $\vec E = \vec{E}_0 e^{-i \omega t}$ with $\vec{E}_0$ a constant electric field and $\sigma$ a tensor.
To compute the conductivity in a holographic framework we turn on a source for the current $\vec{J}$ dual to the gauge field in the bulk. Using the holographic dictionary \cite{Gubser:1998bc,Witten:1998qj}, when the metric on the boundary theory is Minkowski, this current is given by
\begin{equation}
\label{E:JfromA}
\langle J_i \rangle = \frac{\delta S}{\delta A_i^{(0)}} = (d-2) \, {A}_{i}^{(d-2)} \;,
\end{equation}
where $S$ is the on-shell action including possible counter terms and the $A_i^{(n)}$ are given by a near boundary series expansion of the gauge field $\vec{A}$: $A_{i} = \sum_{n=0}^{\infty} A_{i}^{(n)} z^n$. The indices $i,\,j,\,k$ run from $1$ to $d-1$.
The leading coefficient $A_i^{(0)}$ plays the role of the source term.

The bulk equations of motion couple linear fluctuations of the complexified gauge field $A_i$ to other complexified vector fields. In the probe limit these are $\varphi_{ti}$ and $\varphi^*_{ti}$.
This coupling between the vector modes implies that turning on an electric field will source not only the current $\vec{J}$ but also the fluxes $K_i = \langle \cO_{ti} \rangle$ and $K^*_i = \langle \cO^*_{ti} \rangle$.
More explicitly,
\begin{equation}
\vec K = \tau \cdot \vec E
\end{equation}
with $\tau$ a rank two tensor, and
there is a similar equation relating $\vec K^*$ to $\vec E$. Moreover, the source terms for $\langle \cO_{ti} \rangle$ and $\langle \cO^*_{ti} \rangle$ will source not only the fluxes $\vec K$ and $\vec K^*$ but also the current $\vec J$. Thus, in addition to the conductivity and the two aforementioned $(d-1) \times (d-1)$ tensors, we have six additional $(d-1)\times (d-1)$ tensors describing the response of $\vec K$, $\vec K^*$ and $\vec J$ to the source terms for $\langle \cO_{ti} \rangle$ and $\langle \cO^*_{ti} \rangle$.

In what follows we will focus on $d=3$. To compute the frequency dependent conductivity in the $x$-direction, we consider linear, time dependent fluctuations of the complexified fields $A_x$, $\varphi_{ty}$, $\varphi_{ty}^*$, $\varphi_{zy}$ and $\varphi_{zy}^*$. 
At linear order, the coupled algebro-differential equations for the $e^{-i \omega t}$ component of these fluctuations are
\begin{subequations}
\label{E:EOMlinear}
\begin{align}
\label{E:EOMAx}
0 &= A_x'' + \frac{f'}f \, A_x' + \frac{\omega^2}{f^2} \, A_x + \frac{q \psi}{2f^2} \, \big[ (\omega - 2q \phi) \varphi_{ty}^* - (\omega + 2q\phi) \varphi_{ty} \big] \nonumber \\
&\quad - \frac{iq\psi}2 \, \big( {\varphi_{zy}^*}' - \varphi_{zy}' \big) + \frac{iq}{2f} \, (\psi' f - \psi f') \big( \varphi_{zy}^* - \varphi_{zy} \big) \ , \\
\label{E:EOMty}
0 &= \varphi_{ty}'' + \frac2z \, \varphi_{ty}' - \frac{2f + m^2L^2}{z^2 f} \, \varphi_{ty} + L^2 \, \frac{q\omega + 2q^2 \phi}{4z^2 f} \, \psi A_x 
+ \frac i2\, \big[ 2(\omega + q \phi) \varphi_{zy}' + q \phi' \varphi_{zy} \big] \ , \\
\label{E:EOMRzy}
0 &= \big[ (\omega + q\phi)^2 z^2 - m^2L^2 f \big] \, \varphi_{zy} + \frac i4\, L^2 q f \psi A_x' + \frac i2 L^2 q f \psi' A_x \nonumber \\
&\quad - i(\omega + q\phi) z^2 \varphi_{ty}' - \frac i2 \, \big[ 4(\omega + q \phi)z + q \phi'z^2 \big] \, \varphi_{ty} \ .
\end{align}
\end{subequations}
In all but the first equation we expressed $f'$ and $f''$ in terms of $f$.
Notice that since $\varphi_{ty}$, $\varphi_{ty}^*$, $\varphi_{zy}$, $\varphi_{zy}^*$ have to be treated as independent modes, the equations for the starred modes are not the complex conjugate of the unstarred ones---rather they are obtained by complex conjugation and the additional transformation $\omega \to -\omega$. The functions $\varphi_{zy}$ and $\varphi_{zy}^*$ can be eliminated from the first
two equations using (\ref{E:EOMRzy}), leaving three coupled differential equations for $A_x$, $\varphi_{ty}$ and $\varphi_{ty}^*$. Since $A_y$ does not couple to this set of fluctuations, we can conclude that $\sigma_{xy}=0$, implying that there is no Hall conductivity.

The boundary conditions we impose on \eqref{E:EOMlinear} are as follows.  Near the black hole horizon, $A_x$, $\varphi_{ty}$ and $\varphi_{ty}^*$ have the behavior
\begin{equation}
(1-z)^{\pm i \omega / 3} \;.
\label{nearhorizon}
\end{equation}
Requiring that the near-horizon modes of the gauge field and spin two field are falling into the horizon implies that we must choose the minus sign in \eqref{nearhorizon}.
Equivalently, when computing the retarded Green's function for the charge current we should impose causal boundary conditions on the equations of motion \eqref{E:EOMlinear}.
Although with these boundary conditions, $\varphi_{zy} \sim \varphi_{zy}^* \sim (1-z)^{-i \omega / 3 - 1}$, the Lorentz invariant quantity $\varphi_{\mu\nu}^* \varphi^{\mu\nu}$ is finite at $z=z_h$.

\begin{figure}[hbt]
\begin{center}
\includegraphics[height=8cm]{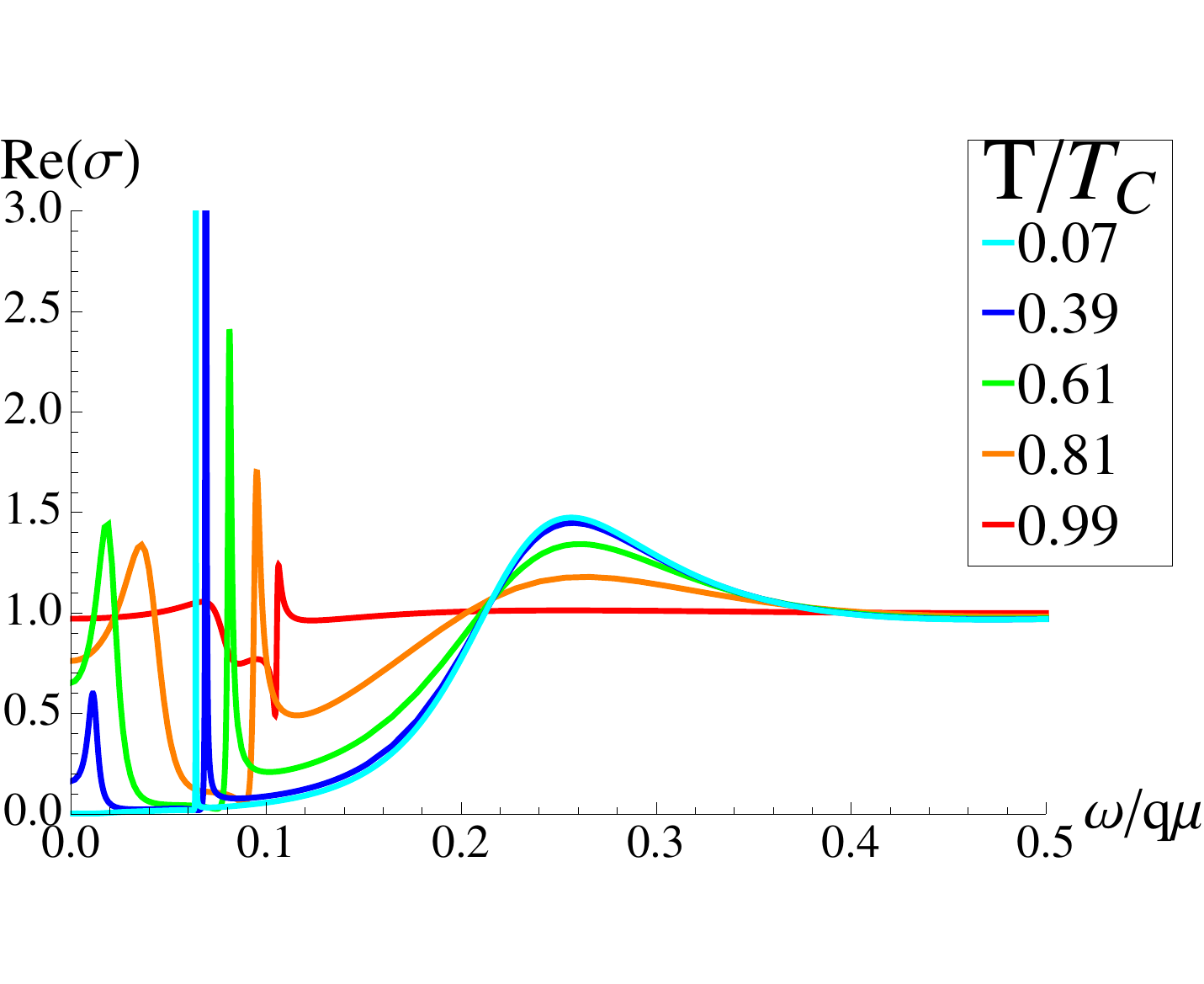}
\caption{\label{F:conductivity} (Color online) The real part of the conductivity as a function of frequency for a $d$-wave condensate of conformal dimension $\Delta=4$. As the temperature is decreased one observes a spike in the conductivity indicative of a bound state. This spike is localized at smaller values of $\omega$ as the temperature is lowered. A second spike in the conductivity appears to vanish as the temperature is decreased.}
\end{center}
\end{figure}

In principle, given the boundary conditions (\ref{nearhorizon}) we could integrate the differential equations for $A_x$, $\varphi_{ty}$ and $\varphi_{ty}^*$ numerically out to the boundary, located at $z=0$, and look for solutions where the leading non-normalizable term in the near boundary series expansion of $\varphi_{ty}$ and $\varphi_{ty}^*$ vanishes.  From the relation between the leading and subleading terms of $A_x$, we can read off the conductivity in the $x$-direction. In practice, since the response of $\varphi_{ty}$ and $\varphi_{ty}^*$ to the electric field is linear, as is the response of $A_x$, $\varphi_{ty}$ and $\varphi_{ty}^*$ to the source terms for $\varphi_{ty}$ and $\varphi_{ty}^*$,
we do not need a shooting algorithm to determine $\sigma$.
Instead, we can directly read off the $xx$ component of the conductivity from the $3\times 3$ matrix relating the expectation values of the boundary fields to the sources. The numerical value of the conductivity for a $\Delta=4$ condensate can be found in figure \ref{F:conductivity}.

It turns out that the (linear) conductivity is isotropic, in spite of the presence of a $d$-wave condensate.
Indeed, to get the conductivity in the $y$ direction, we look at the effect of a $\pi/2$ 
rotation of the condensate in equations (\ref{E:EOMlinear}).
Such a 
rotation flips the sign of $\psi$. By inspection of the equations, 
this is equivalent to flipping the sign of $A_x$.
Since flipping the sign of $A_x$ is equivalent to flipping the sign of both the electric field (the $A^{(0)}$ term) and the current (the $A^{(1)}$ term), the conductivity will be unaffected by such a change of sign. We conclude that the conductivity is proportional to the identity matrix.

We note that the isotropy of the conductivity is only true at linear order. One can check that the non-linear  version of \eqref{E:EOMlinear} does depend on the angle between the electric field and the condensate, so the non-linear response of the material to an electric field is anisotropic.

\section{Fermions}
\label{S:fermions}

In this section, we compute a fermion spectral function in a holographic
superconducting background.
Our calculation is motivated by angle resolved photoemission spectroscopy (ARPES) experiments for $d$-wave superconductors.
ARPES experiments reveal a great deal of information about the various phases of the cuprates and other superconductors.  In these experiments, high energy photons striking the surface of a sample liberate electrons.  The intensity of the produced electrons
is a direct measure of the spectral function of electrons in the sample. Through this virtually direct method of measuring the electron spectral function, ARPES experiments provide an arena where experimental and theoretical results can be compared. See for example \cite{Damascelli:2003bi,Campuzano:2002,Zhou:2006} for reviews of ARPES experiments.

\subsection{The fermionic equations of motion}

Consider a fermionic operator $\cO_\Psi$ in the boundary theory dual to a bulk spinor $\Psi$
whose action is
\begin{equation}
\label{E:ActionSF}
S_\Psi = - \int d^{d+1}x \sqrt{-g} \, {\mathcal L}_\Psi + S_{\rm bnd} \ .
\end{equation}
The Dirac Lagrangian has the standard form
\begin{equation}
\label{E:FermionSimple}
\mathcal{L}_{\Psi} = i \overline{\Psi} \left(\Gamma^{\mu}D_{\mu} - m_\zeta \right) \Psi \;,
\end{equation}
and the sign in \eqref{E:ActionSF} has been chosen to ensure that the spectral function is positive.
The covariant derivative in \eqref{E:FermionSimple} is given by
\begin{equation}
D_{\mu} = \partial_{\mu} + \frac{1}{4} \omega_{\mu,\underline{\lambda\sigma}} \Gamma^{\underline{\lambda\sigma}} -i q_\zeta A_{\mu}
\end{equation}
with $\omega$ the spin connection and $\Psi$ a Dirac spinor of charge $q_\zeta$. The bulk Gamma matrices $\Gamma^{\mu}$ satisfy the Clifford algebra $\{\Gamma^{\mu},\Gamma^{\nu}\} = 2 g^{\mu\nu}$ and $\Gamma^{\mu\nu} \equiv \Gamma^{[\mu}\Gamma^{\nu]}$.
Vielbein indices are underlined. The dimension of $\mathcal{O}_\Psi$ is $\Delta_\Psi$ and is related to the mass of the bulk fermion $m_\zeta$ through
\be
\label{E:Deltafermion}
\Delta_\Psi = \frac d2 \pm m_\zeta L \;.
\ee
The sign in \eqref{E:Deltafermion} depends on the quantization scheme of the boundary theory and must be compatible with the unitarity bound $\Delta_\Psi \geq (d-1)/2$ \cite{Minwalla:1997ka}.
Because our spacetime has a boundary located at $z=0$, we entertain the possibility of a boundary action $S_{\rm bnd}$.  We will discuss the form of $S_{\rm bnd}$ once we have analyzed the near boundary behavior of $\Psi$.

The action (\ref{E:FermionSimple}), while familiar and natural, won't generate interesting dynamics. The fermions in \eqref{E:FermionSimple} are coupled only to the gauge field and the metric and are not directly affected by the spin of the condensing field. Indeed, given the similarity between our condensing spin two field and the Abelian Higgs model of \cite{Hartnoll:2008vx},  the behavior of the fermion correlator in the condensed phase
will be identical to that of the $s$-wave superconductor studied in \cite{Faulkner:2009am,Gubser:2009dt}.

Thinking in the language of effective field theory,
to take into account the effect of the $d$-wave condensate we should add an interaction term to the fermion Lagrangian of suitably low dimension. For definiteness, in what follows we will work with $d=3$. If we consider only interaction terms of mass dimension smaller than six, then we find three possible interaction terms:%
\footnote{%
There are other terms of dimension smaller than six that one can write down in the action. These are: $i|\varphi|^2 \overline\Psi (c + ic\Gamma^5) \Psi$, $\varphi^* \overline{\Psi^c} (c + c\Gamma^5)\Psi$ + h.c., $\varphi^{*2} \overline{\Psi^c} (c + c\Gamma^5)\Psi$ + h.c., $\varphi_\mu^* \overline{\Psi^c} \Gamma^5 \Gamma^\mu \Psi$ + h.c. where all the coupling constants are denoted by $c$.
However, since our ansatz (\ref{ansatz}) satisfies the constraints $\varphi = \varphi_\mu = 0$ and
since $\overline{\Psi^c} \Gamma^\mu \Psi = 0$ due to Fermi
statistics, none of these contribute to the EOM's of the fermion.}
\bea
\label{E:Fterms}
& i|\varphi_{\mu\nu}|^2 \overline{\Psi} \left(c_1 + i c_2 \Gamma^5 \right) \Psi \;, \\
& \varphi_{\mu\nu}^*\varphi^{*\mu\nu} \, \overline{\Psi^c} \left(c_3+c_4\Gamma^5\right) \Psi + \text{h.c.} \;, \\
& \eta^* \varphi^*_{\mu\nu} \overline{\Psi^c}\Gamma^{\mu}D^{\nu} \Psi + \text{h.c.} \;,
\eea
where $\Psi^c \equiv C \Gamma^\ut \Psi^*$ and the charge conjugation matrix $C$ is defined by $C \Gamma^\mu C^{-1} = - \Gamma^{\mu\trans}$.
The first term contributes to an effective mass for the fermion, without any angular dependence related to the $d$-wave condensate. Since we do not expect any interesting physics to arise from it we will not consider it further. The second and third terms are not mutually compatible: the second term requires that the charge of the fermion, $q_{\zeta}$, equal the charge of the condensate, $q$, while the third term requires $2q_\zeta = q$. The second term in \eqref{E:Fterms} will produce an $s$-wave gap similar to the one studied in \cite{Faulkner:2009am} in a slightly different context.

In what follows we will focus on the last term in (\ref{E:Fterms}) since it has the most interesting effect on the fermion correlator.
Thus, the bulk fermion Lagrangian we consider is given by
\begin{equation}
\label{E:FermionAction}
\mathcal{L}_{\Psi} = i \overline{\Psi} \left(\Gamma^{\mu}D_{\mu} - m_\zeta \right) \Psi
+\eta^* \varphi^*_{\mu\nu} \overline{\Psi^c}\Gamma^{\mu}D^{\nu} \Psi
-\eta \overline{\Psi} \Gamma^{\mu} D^{\nu} \left(\varphi_{\mu\nu}\Psi^c\right) \;.
\end{equation}
Since the phase of $\eta$ can be changed by a redefinition of the fermion field, we will set $\eta \geq 0$ in the rest of this discussion. The Dirac equation following from \eqref{E:FermionAction} is
\be
\label{Dirac equation}
0 = \big( \Gamma^\mu D_\mu - m_\zeta \big) \Psi + 2i\eta \varphi_{\mu\nu} \Gamma^\mu D^\nu \Psi^c + i \eta \varphi_\mu \Gamma^\mu \Psi^c \;.
\ee
It will be convenient to rewrite the equations of motion in terms of a rescaled four-component spinor $\zeta = (-g\cdot g^{zz})^{1/4} \Psi$ which we will further decompose into two two-component spinors $\zeta = (\zeta_1, \zeta_2)$.
Since the Dirac equation involves both $\Psi$ and $\Psi^c$ then, in Fourier space, it will couple modes with wavevector $(\omega,\vec k)$ to modes with wavevector $(-\omega, - \vec k)$. Thus, we look for solutions to \eqref{Dirac equation} which are of the form
\be
\zeta = e^{-i\omega t + i \vec k\cdot \vec x} \, \zeta^{(\omega,\vec k)}(z) + e^{i\omega t - i \vec k\cdot \vec x} \, \zeta^{(-\omega,-\vec k)}(z) \;,
\ee
where the notation $\zeta^{(\omega,\vec k)}$ implies a single Fourier mode.
We make the following choice of (real) bulk gamma matrices:
\be
\Gamma^{\ut} = \mat{ - i\sigma_2 & 0 \\ 0 & i \sigma_2 } \qquad
\Gamma^{\uz} = \mat{ \sigma_3 & 0 \\ 0 & \sigma_3 } \qquad
\Gamma^{\ux} = \mat{ \sigma_1 & 0 \\ 0 & \sigma_1 } \qquad
\Gamma^{\uy} = \mat{ 0 & -i\sigma_2 \\ i\sigma_2 & 0 }
\label{E:gammas}
\ee
and use $C = \Gamma^{\ut}$. The Pauli matrices are denoted $\sigma_i$. The boundary gamma matrices
$\gamma^{m}$ are defined by the action of the Lorentz generators $i \Gamma^{\ulambda\usigma}$ on the positive eigenspace of $\Gamma^{\uz}$.
We find that $\gamma^t = -i\sigma_2$, $\gamma^x = \sigma_1$, $\gamma^y = -\sigma_3$ and $C = - \gamma^t$.
With these definitions, the Dirac equation \eqref{Dirac equation} reduces to
\bea
\label{E:Deqns}
0 &= D_{(1)} \, \zeta^{(\omega,k_x)}_1 + 2\eta (g^{xx})^{3/2} k_x \big[ \varphi_{xx-yy} \sigma_1 \zeta^{(-\omega,-k_x)*}_1 - i \varphi_{xy} \sigma_2 \zeta^{(-\omega,-k_x)*}_2 \big] \\
0 &= D_{(2)} \, \zeta^{(\omega,k_x)}_2 + 2\eta (g^{xx})^{3/2} k_x \big[ \varphi_{xx-yy} \sigma_1 \zeta^{(-\omega,-k_x)*}_2 + i \varphi_{xy} \sigma_2 \zeta^{(-\omega,-k_x)*}_1 \big]
\eea
where we have aligned the momentum vector $\vec{k}$ along the $x$ axis. The solution for arbitrary $\vec k$ can be recovered by using (\ref{condensate rotation}). The differential operators $D_{(\alpha)}$ are given by
\be
D_{(\alpha)} = \sqrt{g^{zz}} \, \sigma_3 \partial_z - m_\zeta +(-1)^{\alpha} (\omega + q_\zeta A_t) \sqrt{-g^{tt}}\, \sigma_2 + ik_x \sqrt{g^{xx}} \, \sigma_1 \;.
\ee
The equations for $\zeta^{(-\omega,-k_x)}$ and 
for $\zeta^{(\omega,k_x)*}$ can be obtained from \eqref{E:Deqns} by an appropriate sign flip or complex conjugation.

We are interested in the retarded Green's function for the boundary
fermion operator $\mathcal{O}_{\Psi}$ in the condensed phase at non-zero temperature.
The prescription we use to study this fermion correlation function closely follows that of \cite{Iqbal:2009fd,Liu:2009dm,Faulkner:2009wj,Cubrovic:2009ye} which is based on the
work of \cite{Son:2002sd}.  In what follows we will summarize this prescription leaving a more detailed review to Appendix \ref{app:FGreens}.
Consider a solution to the Dirac equation \eqref{E:Deqns} with infalling boundary conditions at the black hole horizon. We denote the coefficients of a near boundary expansion of the fermion fields which solve \eqref{E:Deqns} as $S$ and $R$ where
\be
\zeta_\alpha = \mat{ O(z) \\ (\sigma_1 S)_\alpha } z^{-m_\zeta L} + \mat{ R_\alpha \\ O(z)} z^{m_\zeta L} \;.
\label{RSdef}
\ee
The index $\alpha = 1,2$ is a boundary spinor index, and the expansion above applies to both $\zeta^{(\omega, \vec k)}$ and $\zeta^{(- \omega, - \vec k)}$. Here $S$ is identified with the source for the operator $\cO_\Psi$, while $R = \langle \cO_\Psi \rangle$.
Since the Dirac equation \eqref{Dirac equation} is linear, $R^{(\omega,\vec k)}$ will be linearly related to $S^{(\omega,\vec k)}$ and $S^{(-\omega,-\vec k)\,c}$:
\be
R_\alpha^{(\omega,\vec k)} = \cM\du{\alpha}{\beta} S_\beta^{(\omega, \vec k)} + \widetilde{\cM}\du{\alpha}{\beta} S_\beta^{(-\omega,- \vec k)\,c} \;.
\label{bulklinrespond}
\ee
The prescription to compute the retarded Green's function is then \cite{Iqbal:2009fd}
\be
G_R(\omega,\vec k) = - i \cM \gamma^t \;.
\label{GRanswer}
\ee
Typically the spectral function is defined to be the matrix quantity
$
\rho(\omega, \vec k) = \big( G_R(\omega, \vec k) -  G_R^\dagger(\omega, \vec k) \big) /2i
$.
With a little abuse of terminology, we define the spectral function as
\be
\label{E:rhodef}
\rho(\omega, \vec k) \equiv \Tr \Im G_R(\omega, \vec k)
\ee
where the trace is over spinor indices.

\subsection{The spectral function for fermion operators}

We obtained the spectral function \eqref{E:rhodef} numerically using \eqref{GRanswer}.
After obtaining a numerical solution for the condensate (described in section \ref{S:dwaveCondensate}) we solve the Dirac equation \eqref{E:Deqns} and numerically extract the near boundary components $S$ and $R$ defined in \eqref{RSdef}. In what follows we focused on $m_\zeta = 0$, leaving the exploration of the mass parameter to future work.

Consider first a configuration in which the coupling of the fermions to the spin two field vanishes, $\eta = 0$. As discussed earlier, since the fermions are not coupled to the condensate, the resulting Dirac equation is identical to the one studied in \cite{Faulkner:2009am,Gubser:2009dt}.
For configurations in which the near horizon geometry is asymptotically AdS and at finite $q$ and zero temperature, refs.~\cite{Faulkner:2009am,Gubser:2009dt} found that the spectral function for the fermions has continuous support inside the lightcone, $|\omega| \geq c_{IR} |\vec k|$, where $c_{IR}$ is the speed of light in the infrared.
For $|\omega| < c_{IR} |\vec k|$ the spectral function has delta-function support along codimension-one surfaces, which are in one-to-one correspondence with zero-modes of $\zeta$.
At $T=0$ one identifies
the momentum $|\vec k|=k_F^{(a)}$ where a mode crosses the $\omega=0$ axis with
a Fermi momentum in the boundary theory.
As expected for an $s$-wave configuration, the spectral function is rotationally symmetric.

\begin{figure}[t]
\begin{center}
\includegraphics[width=15cm]{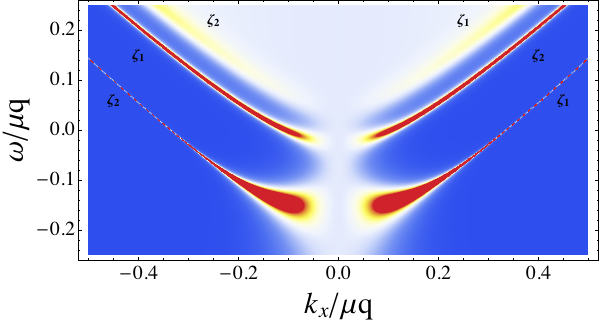}
\caption{(Color online) The fermion spectral function for decoupled ($\eta = 0$) massless fermions at $T=0.66 T_c$. Red implies a large value of the spectral function and blue represents a value closer to zero. The red bands in the spectral function are associated with quasi-normal modes of $\zeta_1$ and $\zeta_2$. As described in the text, if the dispersion relation of $\zeta_1$ is $\omega = E(k_x)$ then that of $\zeta_2$ is $\omega = E(-k_x)$.
\label{F:eta0}}
\end{center}
\end{figure}

As opposed to \cite{Faulkner:2009am,Gubser:2009dt} , we work at non-zero temperature and in the probe limit where the geometry is unaffected by the matter fields. As we approach $T=0$ the geometry approaches empty AdS and, in particular, its speed of light in the infrared is one.\footnote{%
We once again emphasize that when $m^2>0$ the analysis carried out in \cite{Gubser:2009cg,Horowitz:2009ij} implies that one can not consistently take $q \to \infty$ and then $T \to 0$. Therefore, in this work we consider only $T>0$.}
We conclude that in the probe limit, and at low but non-zero temperature, there should be an approximate IR lightcone, given by $|\omega| \geq |\vec k|$, in which the spectral function has continuous support. Also, the zero-modes which will be observed at zero temperature should broaden into quasi-normal modes.
These broadened quasi-normal modes can be seen in figure \ref{F:eta0}.

Still working with $\eta = 0$ we focus, without loss of generality, on the $k_y=0$ plane. The equations of motion for
$\zeta_1$ and $\zeta_2$  decouple and we can separately discuss the quasi-normal modes of $\zeta_1$ and $\zeta_2$. These quasi-normal modes appear as peaks in the components $G_{R,11}(\omega,k_x)$ and $G_{R,22}(\omega,k_x)$ respectively. Consider a single quasi-normal mode of, say, $\zeta_1$. We define the dispersion relation of this quasi-normal mode as the location of the maxima of $G_{R,11}$ and denote it by $\omega=E_1(k_x)$. Generically there will be multiple quasi-normal modes so we add an index $a$ to the dispersion relation $\omega=E_1^{(a)}(k_x)$. Similarly, $\zeta_2$ will have quasi-normal modes with dispersion relation $\omega=E_2^{(a)}(k_x)$.
For $\eta = 0$ the equation of motion for $\zeta_2$ can be obtained from the equation of motion for $\zeta_1$ by making the substitution $\zeta_1 \to \sigma_3 \zeta_2$ and $k_x \to -k_x$. Therefore, for each quasi-normal mode of $\zeta_1$ there is a corresponding quasi-normal mode of $\zeta_2$ and vice versa, such that $E_2^{(a)}(k_x) = E_1^{(a)}(-k_x)$. See figure \ref{F:eta0}. We note in passing that the quasi-normal modes of $\zeta_i^*$ have dispersion relation $-\omega = E_i^{(a)}(-k_x)$.

\begin{figure}[t]
\includegraphics[height=5.3cm]{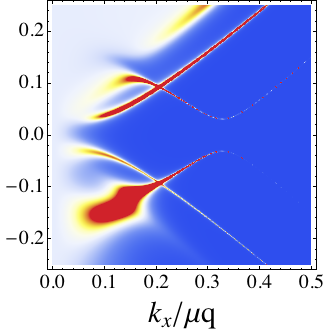}
\includegraphics[height=5.3cm]{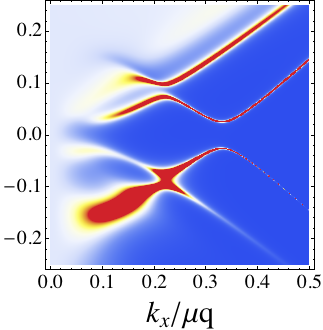}
\includegraphics[height=5.3cm]{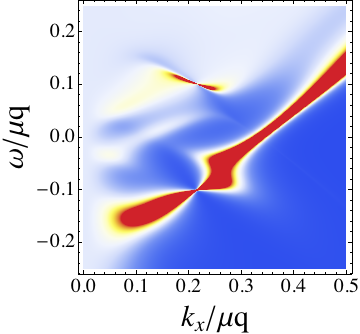}
\caption{(Color online) The spectral function for the fermions
evaluated for $\eta=0.5$ and $T=0.66 T_c$. The red bands corresponds
to a large value of the spectral function and the blue regions to
values which are closer to zero. The left panel corresponds to the
anti-node, the central panel to a $22.5$ degree angle from the node
and the value of the spectral function at the node is shown in the
right panel.
\label{F:bands}}
\end{figure}

A non-vanishing, but small, $\eta/L$ in \eqref{E:Deqns}  couples the quasi-normal modes of $\zeta^{(\omega,k_x)}$ and $\zeta^{(-\omega, -k_x)*}$ with an angle dependent strength, 
$\theta$ being the angle between the condensate and the momentum vector. We define $\theta=0$ as the angle for which $\varphi_{xy}\neq 0$, $\varphi_{xx-yy}=0$ and $\vec k = (k_x, 0)$. The spectral function for this configuration is depicted in the left panel of  figure \ref{F:bands}. At this angle $\zeta_1^{(\omega,k_x)}$ is coupled to $\zeta_2^{(-\omega,-k_x)*}$. Once the coupled modes intersect, eigenvalue repulsion will replace the intersection by a gap. This gap-generating feature of (\ref{E:FermionAction}) is similar to the mechanism
described in \cite{Faulkner:2009am}, where a gap was generated in an $s$-wave superconductor by adding a Majorana-like coupling to the fermion action. The set of lifted degeneracies includes all those located at $\omega = 0$.

Next, consider a configuration at $\theta = \pi/4$ where $\varphi_{xx-yy} \neq 0$, $\varphi_{xy} = 0$ and $\vec k = (k_x, 0)$ (this configuration is equivalent to one with $k_x = k_y$ and $\varphi_{xx-yy} = 0$).
The spectral function for this configuration appears on the right panel of figure \ref{F:bands}. We find that $\zeta_1^{(\omega,k_x)}$ is coupled to $\zeta_1^{(-\omega,-k_x)*}$ and $\zeta_2^{(\omega,k_x)}$ is coupled to $\zeta_2^{(-\omega,-k_x)*}$.
The set of lifted degeneracies is distinct from the one at $\theta = 0$.
We call $\theta = 0$ the anti-nodal direction and $\theta = \pi/4$ the nodal direction. A node is a point on the $\omega=0$ axis where two quasi-normal modes intersect. At $\theta = \pi/4$, $\zeta_1^{(\omega,k_x)}$ is not coupled to $\zeta_2^{(-\omega,-k_x)*}$ and since only $\zeta_1^{(\omega,k_x)}$ and $\zeta_2^{(-\omega,-k_x)*}$ have dispersion curves that intersect along the $\omega=0$ axis, there will be no  associated gap at such an angle. At intermediate angles $\theta \neq 0$, $\pi/4$ all the degeneracies are lifted.%
\footnote{To be more precise, we should write $\theta \neq 0 + n \, \frac\pi2$, $\theta \neq \frac\pi4 + n \, \frac\pi2$, because the spectral function is invariant under a $\pi/2$ rotation:
The spin two field $\varphi_{\mu\nu}$ changes sign under a rotation of $\pi/2$, but this sign change is equivalent to changing the sign of $\eta$ which in turn can be absorbed in a redefinition of the fermion field.}
This lifting is depicted in the central panel of figure \ref{F:bands}.

We note that once $\eta/L$ becomes too large our perturbation-theory based analysis breaks down and we find that a gap is generated at all angles. The critical value of $\eta$ for which this happens depends on $T/q_\zeta \mu$, $\Delta$ and $m_\zeta L$.

There are other interesting features that come out of the analysis of the spectral function and that, surprisingly, compare well with properties of $d$-wave superconductors. For instance the precise angular dependence of the gap is well fit by $| \cos k_x - \cos k_y |$, the nodes broaden into Fermi arcs in a particular temperature range, and Dirac cones are observed with Fermi velocities whose ratio can be tuned by dialing $\eta$. This analysis is performed in the companion letter \cite{Benini:2010qc}.

\subsection{The physics of bulk fermions}

\begin{figure}
\begin{center}
a) \includegraphics[width=1in]{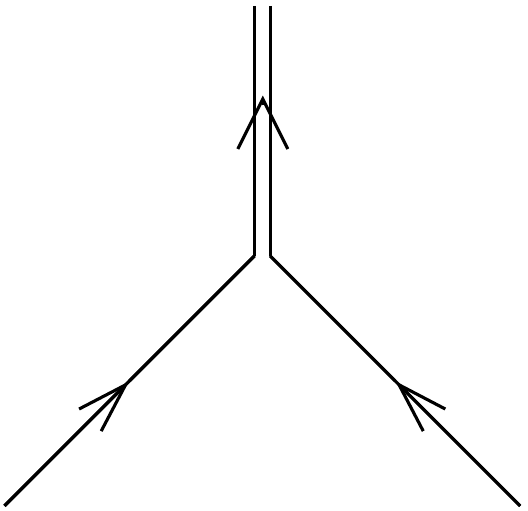}
\hskip 1in
b) \includegraphics[width=1in]{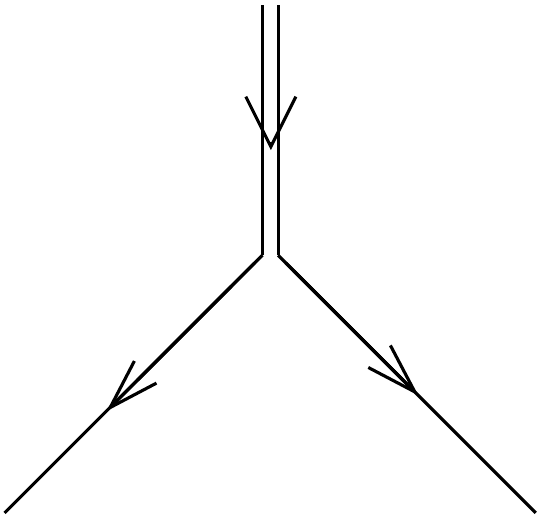}
\end{center}
\caption{
The Feynman diagrams associated with the interaction terms 
between the fermions and the spin two field in
 (\ref{E:Fterms}):
(a) 
$\eta^* \varphi^*_{\mu\nu} \overline{\Psi^c}\Gamma^{\mu}D^{\nu} \Psi$; 
(b) 
$\eta \overline{\Psi} \Gamma^{\mu} D^{\nu} \left(\varphi_{\mu\nu}\Psi^c\right)$. 
\label{fig:vertices}
}
\end{figure}

The bulk fermions are charged under a $U(1)$ field and we can think of them as electrons and positrons. In the presence of a spin two condensate, the vertex
$\varphi^*_{\mu\nu} \overline {\Psi^c} \Gamma^\mu D^\nu \Psi$ and its Hermitian conjugate lead to electron positron mixing.
See figure \ref{fig:vertices}.
These interaction vertices have an interesting spin structure which we now elaborate.

Consider the Dirac equation, $D_{(\alpha)} \zeta_\alpha = 0$, in a limit (and region of space-time) where $k_x = A_t = 0$ and where the $z$-derivative can be ignored. In this limit, we can find approximate solutions to the Dirac equation of the form
\be
\zeta_1^{(\omega,0)} \approx \zeta_2^{(\omega,0)*} \approx \mat{1 \\ -i} \;,\qquad\qquad
\zeta_2^{(\omega, 0)} \approx \zeta_1^{(\omega,0)*} \approx \mat{1 \\ i} \;,
\ee
where $\omega = m_\zeta / \sqrt{-g^{tt}}$.  The Lorentz generator of rotations in the $xz$-plane is given by
\be
\Sigma^{\uy} = i \Gamma^{\uz} \Gamma^{\ux} = \mat{ - \sigma_2 & 0 \\ 0 & -\sigma_2 } \;.
\ee
Thus, $\zeta_1^{(\omega,0)}$ has spin-up in the $y$-direction while $\zeta_2^{(\omega,0)}$ has spin down in the $y$ direction. In other words, in an approximate rest frame for the fermions, the quantum field $\zeta_1^{(\omega,0)}$ annihilates bulk spin up electrons in the $y$ direction (for $\omega>0$) or creates bulk spin down positrons (for $\omega<0$), while $\zeta_2^{(\omega,0)}$ annihilates bulk spin down electrons or creates bulk spin up positrons.

To gain further physical insight into the existence of these quasi-normal modes, consider a semiclassical picture of an electron with positive electric charge $q_\zeta >0$, positive spin in the $y$ direction, and $k_x > 0$. Such an electron is represented by $\zeta_1^{(\omega, k_x)}$. This electron feels both a gravitational attraction to the black hole and an electrostatic repulsion from the electric field $E_z < 0$.  If the electric field becomes strong enough, a bound state appears in the potential for $\zeta_1^{(\omega, k_x)}$.

If we boost to the rest frame of the electron, we observe  a magnetic field $B_y<0$ in addition to the electric field $E_z$.  This magnetic field will act to lower the energy of the electron because of the coupling with its spin.
In other words, there is an additional shift to the energy level $E_1(k_x)$ of the bound state because of spin-orbit coupling. For the corresponding electron with spin down ($\zeta_2^{(\omega, k_x)}$) the energy $E_2(k_x)$ is increased.  If the potential, before considering spin-orbit effects, is not deep enough, the increase will lead to the loss of the bound state.
However, as the electric field is increased, the potential well supports more and more bound states, and we see more and more alternating bands of $\zeta_1$ and $\zeta_2$ quasinormal modes. For large mass fermions, we expect the spin-orbit coupling to be a small effect, smaller than the spacing between the bound states in the potential, ensuring that the bands are alternating.  In the small mass limit, our non-relativistic quantum mechanics intuition fails, but the numeric calculations of the previous section indicate that the bands continue to be alternating.

The Fermi sea in the bulk consists of a series of concentric disks in the $xy$-plane in $k$-space, one disk per band of normal modes.  Each disk possesses a Fermi surface at $\omega=0$.
In the limit in which $k_x$ is small and the rest frame calculation of the spin is a good approximation,
as we move around one of these Fermi surfaces in $k$-space, the spin of the electron rotates in the $xy$ plane, always pointing tangent to the surface.
In particular, for $k_x < 0$ and $k_y=0$, the role of $\zeta_1$ and $\zeta_2$ are switched because the sign of the magnetic field is reversed if we boost in the opposite direction.

From a bulk point of view, the coupling between the spin two field and  the fermions produces new quasiparticles in the presence of a spin two condensate.
In $k$-space, at $\theta=0$, electrons can mix with positrons of the same spin on opposite sides of the Fermi sea---$\zeta_1$ with $\zeta_2^*$ or $\zeta_2$ with $\zeta_1^*$.
However, at $\theta=\pi/4$ electrons mix with positrons with opposite spin on opposite sides of the bulk Fermi sea---$\zeta_1$ with $\zeta_1^*$ or $\zeta_2$ with $\zeta_2^*$.  This type of new quasiparticle with indefinite charge also appears in BCS theory.

The spin structure of the bulk Fermi surface is crucial for the angle dependence of the gap.  Naively, in the absence of a magnetic field, one would expect fermions with different spin to have the same energy at the Fermi surface.  Because of the spin-orbit coupling, we find in fact that the energy levels are split.  The dispersion curves for $\zeta_1$ and $\zeta_1^*$  do not cross at the Fermi surface while the curves for $\zeta_1$ and $\zeta_2^*$ do.  Thus we find a node at $\theta = \pi/4$ and a gap at $\theta=0$.

\paragraph{Do the fermions back react?} Adding massless fermions to the system with charge $q_\zeta$, we would like to estimate semi-classically the total amount of charge contained in the Fermi sea.  We restrict to $d=3$ for this estimate, and we work locally, assuming the curvatures are small and space looks locally flat.
With a potential of the form (\ref{E:normalphase}), the Fermi energy will be
\begin{equation}
E_F = q_\zeta A_t \sqrt{-g^{tt}} = \frac{z}{L \sqrt{f}} \mu q_\zeta \left( 1 - \frac{z}{z_h} \right) \;.
\end{equation}
For a free massless gas of fermions, $E_F = (3 \pi^2 \rho)^{1/3}$ where $\rho$ is the local number density.
Thus we have
\begin{equation}
\rho = \frac{1}{3 \pi^2} \left[ \frac{z}{L \sqrt{f}} \mu q_\zeta \left( 1- \frac{z}{z_h} \right) \right]^3 \;.
\end{equation}
The number density in the field theory from the fermions can be estimated by performing the integral
\be
n = \int_0^{z_h} \frac{ \rho L^3}{z^3 \sqrt{f} } \, dz = (\sqrt{3} \pi-3) \, \frac{\mu^3 q_\zeta^3}{27 \pi^2} \, z_h \;.
\ee
The field theory charge density from $A_t$ on the other hand is $\mu / z_h$.  To be able to ignore the back reaction of the fermions on the gauge field, we need
\begin{equation}
\mu^2 z_h^2 q_\zeta^4 \ll 1 \;.
\end{equation}
For this condition to be met at temperatures of order the phase transition temperature, we need
$\Delta^2 q_\zeta^4 / q^2 \ll 1$.  We have chosen $q_\zeta = q / 2$ in previous sections,
and so $ q  \Delta \ll 1$.

Earlier, we mentioned that the probe limit involves taking $q$ large and one might worry that the limit is incompatible with ignoring the back reaction of the fermions on the gauge field.  In fact, we should be more careful.
Although we don't have an action for the spin two field where we can rigorously take back reaction into account, we can look at the back reaction of the gauge field on the metric for temperatures close to $T_c$.  In this case, we know the fully back reacted solution involves a Reissner-Nordstrom black hole in AdS.
Placing a $1/ 2 \kappa^2$ in front of the gravitational action,
we find that the back reaction modifies the warp factor:
\begin{equation}
f(z) = 1 - \left( \frac{z}{z_h} \right)^3 ( 1 + Q) + \left( \frac{z}{z_h} \right)^4 Q \;,
\end{equation}
where $Q = z_h^2 \kappa^2 \mu^2 / 2 L^2$.
In the normal phase the probe limit condition is thus
\begin{equation}
\kappa z_h \mu \ll L \;.
\end{equation}
Close to $T_c$, using (\ref{E:TcAnalytic}), the inequality reduces to $q \gg m \kappa$.  Thus, if we can tune $m \kappa$ to be suitably small, then there is an intermediate coupling regime
$m \kappa \ll q \ll 1 / m L$ where the probe limit is valid and fermionic back reaction can be ignored.  Whether such a regime can be realized in a stringy construction is another question.

\section{Outlook}

In this work we have discussed the construction of a holographic $d$-wave superconductor. As emphasized several times in the text, our construction is not ideal. There are well-known difficulties in writing down an action for a charged spin two field propagating in a curved spacetime. These difficulties lead to non-hyperbolic and non-causal behavior of the spin two field. In principle these problems could be circumvented when considering the phase diagram of the theory and in studying the fermion correlators in the condensed phase if one works with a very massive spin two field in the bulk (or, equivalently, a spin two operator of large conformal dimension in the boundary). Numerically, it is challenging to construct backgrounds for fields of very large mass and it would be worthwhile to improve on our current results where we studied condensates of dimension less than or equal to six. Alternately, one might try to use analytic techniques, perhaps similar to the ones used in appendix \ref{app:semiclassics} or in \cite{Kraus:2002iv} to construct these solutions.

It would be, of course, much more satisfactory to have an action that produces hyperbolic and causal  equations of motion with the correct number of degrees of freedom. Unfortunately, such an action is not presently known. There are, however, several different directions one could try to pursue. In \cite{Porrati:2009bs} a consistent action for a spin $3/2$ particle propagating  in a constant electro-magnetic field and flat spacetime was constructed. Similar techniques might be used to obtain an improved form of the Argyres-Nappi action \cite{Argyres:1989cu} which would have the correct number of propagating degrees of freedom in dimensions other than 26. A different approach that would lead to a causal action for a spin two particle would be to consider a Kaluza-Klein compactification of Einstein gravity on a manifold with a $U(1)$ isometry. The drawback of such a method is that, generically, such a compactification would lead to an action describing a whole tower of massive spin two fields. A final method which might prove useful would be an expansion of our proposed action in powers of $F_{\mu\nu}$ and its derivatives, requiring that the resulting equations of motion are causal and have the right number of degrees of freedom at each order of the expansion.

Even without a full-fledged causal action for the spin two field, there are certain properties of the condensate that may be probed. In this paper we discussed only solutions where  $\varphi_{xy}$ and $\varphi_{xx-yy}$  are real. By introducing a relative complex phase between these modes, one obtains an exotic $d+id$ order parameter that breaks time reversal invariance. This $d+id$ superconductor will probably have a non-vanishing Hall conductivity (with spontaneous currents along the boundaries of the sample) in the absence of a magnetic field.  Similar phenomena were reported in a holographic context for $p$-wave superconductors in \cite{Gubser:2008wv, Roberts:2008ns}.

In addition to the conductivity, we studied the effect of the spin two condensate on probe fermions.  We saw how the angle dependence of the gap emerged from simple qualitative arguments involving degenerate perturbation theory and spin-orbit coupling.
Given the difficulty of solving the Dirac equation, it would be interesting to see if semi-classical techniques such as the WKB approximation could be used to further elucidate the physics of fermions in these black hole backgrounds.

\section*{Acknowledgments}

We would like to thank Massimo Porrati, Silviu Pufu, Fabio Rocha, Subir Sachdev, Ronny Thomale and Yuji Tachikawa for discussions. FB thanks Adriano Amaricci and Cedric Weber for patient explanations. The work of FB and CH was supported in part by the US NSF under Grants No.\
PHY-0844827 and PHY-0756966.
The work of RR was supported by ERC Grant n.226455 Supersymmetry,
Quantum Gravity and Gauge Fields (Superfields).
The work of AY was supported in part by the US DOE under
Grant No.\ DE-FG02-91ER40671.

\appendix

\section{The spin two Lagrangian}
\label{app:constraints}

It is difficult to write down an action for a massive charged spin two field in a curved spacetime. Most of the difficulties are associated with irregularity of the spin two Lagrangian which leads to a constrained system of equations. In what follows we will begin by reviewing the Fierz-Pauli Lagrangian for a neutral, non-interacting spin two field in Minkowski space \cite{Fierz:1939ix}. Many of the strategies used in this simple case can be abstracted to the more complicated example of interest.

\subsection{Neutral field in flat spacetime}
\label{A:FPEOM}

The EOM's that follow from the Fierz-Pauli Lagrangian (\ref{E:FP}) are
\begin{equation}
E_{\mu\nu} = \varphi_{\mu\nu,\lambda}^{\phantom{\mu\nu,}\lambda} - \varphi_{\mu,\nu} - \varphi_{\nu,\mu}
+ \varphi_{,\mu\nu} - m^2 \varphi_{\mu\nu} - \eta_{\mu\nu} ( \varphi_{,\lambda}^{\phantom{,}\lambda} - {\varphi_{\lambda,}}^\lambda - m^2 \varphi) =0 \;.
\end{equation}
Inspection reveals that the $d+1$ equations $E_{\mu t}=0$ do not involve second order time derivatives.  Thus, the Lagrangian \eqref{E:FP} is irregular and the equations of motion
$E_{\mu t} = 0$ are actually constraints.%
\footnote{For a brief explanation and classification of constrained systems see \eg{} \cite{Weinberg}.}
The existence of constraints is expected since the
symmetric tensor $\varphi_{\mu\nu}$ naively has $d+2$ more degrees of freedom than a
massive spin two particle.
In order to fix the values of the extra $d+2$ fields and their momenta we need $2(d+2)$ constraint equations. The divergence of $E_{\mu\nu}$ involves only first derivatives and yields $d+1$ further constraints:
\begin{equation}
\label{divE}
{E_{\mu\nu,}}^\mu  = m^2 ( \varphi_{,\nu} - \varphi_\nu )=0 \;.
\end{equation}
A penultimate constraint can be obtained by using the trace of the EOM's,
\begin{equation}
\label{trE}
E_\mu^\mu = (d-1)( {\varphi_{\lambda,}}^\lambda - \varphi_{,\lambda}^{\phantom{,}\lambda} ) + d \,  m^2 \varphi =0 \;,
\end{equation}
together with ${E_{\mu\nu,}}^{\mu\nu}=0$ which yields
\begin{equation}
\varphi = 0 \;.
\end{equation}
The last constraint is $\varphi_{,t} = 0$.
In this simple case, the EOM's and constraints can be rewritten in the form
(\ref{E:FPequations}), which is the form of Fierz.

\subsection{Charged field in curved spacetime}

Having tackled the Fierz-Pauli action, we are better prepared to deal with the more involved case of a charged massive spin two field in a curved background. Our treatment follows 
\cite{Fierz:1939ix,Buchbinder:1999ar,Buchbinder:2000fy}.

Let us write down the most general $d+1$ dimensional action, quadratic in the spin two field  and of scaling dimension $d+1$:
\bea
\label{E:ActionInit}
\cL &= - |D_\rho \varphi_{\mu\nu}|^2 + 2|D_\mu \varphi^{\mu\nu}|^2 + |D_\mu \varphi|^2 - \big[ D_\mu \varphi^{*\mu\nu} D_\nu \varphi + \text{h.c.} \big] - m^2 \big( |\varphi_{\mu\nu}|^2 - |\varphi|^2 \big) \\
&\quad + c_1  R_{\mu\nu\rho\lambda} \varphi^{*\mu\rho} \varphi^{\nu\lambda} + c_2 R_{\mu\nu} \varphi^{*\mu\lambda} \varphi^\nu_\lambda + c_3 \big[ \, e^{i \theta} R_{\mu\nu} \varphi^{*\mu\nu} \varphi + \text{h.c.} \big] + c_4 R |\varphi_{\mu\nu}|^2 + c_5 R|\varphi|^2 \\
&\quad + i c_6 q F_{\mu\nu} \varphi^{*\mu\lambda} \varphi^\nu_\lambda \;.
\eea
The $c_i$ are assumed to be real.
The terms in the first line of (\ref{E:ActionInit}) are fixed by requiring that \eqref{E:ActionInit} reduces to \eqref{E:FP} when the metric is Minkowski and when the charge of the spin two field vanishes.
The terms on the second line of (\ref{E:ActionInit}) involve all possible curvature invariants. They vanish on a flat spacetime.  There is only one possible invariant we can construct from $F_{\mu\nu}$ that is quadratic in the $\varphi_{\mu\nu}$ and of scaling dimension $d+1$.

We point out two subtleties in this Lagrangian.
First, we can rule out a contribution of the form
\begin{equation}
i \big[ D_\mu \varphi^{*\mu\nu} D_\nu \varphi - \text{h.c.} \big]
\end{equation}
because in the $q \to 0$ limit such a term would mix the real and imaginary parts of $\varphi_{\mu\nu}$ instead of leading to a Lagrangian for two, free, real spin two fields.
Second, there is a potential ambiguity that involves a combination of the first and second lines:
the expression
\be
\label{E:totalD}
\sqrt{-g} \, \big(- D_\rho \varphi^{*\rho\mu} D^\lambda \varphi_{\lambda\mu}
+ D^\lambda \varphi^{*\rho\mu} D_\rho \varphi_{\lambda\mu} - R_{\mu\nu\lambda\rho} \varphi^{*\mu\lambda} \varphi^{\nu\rho} + R_{\lambda\rho} \varphi^{*\lambda\mu} \varphi^\rho_\mu + i F_{\mu \nu} \varphi^{*\mu \lambda} \varphi^\nu_\lambda \big)
\ee
is a total derivative.%
\footnote{Our convention for the Riemann tensor is that $[\nabla_\mu, \nabla_\nu] V^\lambda = {R^\lambda}_{\rho \mu \nu} V^\rho$.}
We have resolved this ambiguity by setting the $(D_\lambda \varphi^*_{\mu\nu}) (D^\mu \varphi^{\lambda \nu})$ term in the first line to zero and allowing for $c_1$, $c_2$, and $c_6$ in the second and third lines.

We will fix the $c_i$ and $\theta$ by requiring the existence of $2(d+2)$ constraint equations.
We begin by computing the EOM's: $E_{\mu\nu}=0$.
Just as in flat space, the equations $E_{\mu t}=0$ do not involve second derivatives with respect to time and constitute $d+1$ constraints.  Taking the divergence of the EOM's,
$D^\mu E_{\mu\nu}=0$, we find $d+1$ additional constraints.

To find the two remaining constraints, we pursue the same strategy used in flat space: we would like to combine the trace equation $E_\mu^\mu = 0$ with  $D^\mu D^\nu E_{\mu\nu}=0$ in such a way as to eliminate the second derivative terms, leaving at most first derivatives of the fields.
The trace of the EOM's, $E_\mu^\mu=0$, reads
\begin{equation}
\label{trEcurved}
\big[ d \, m^2 + \big(c_3 + c_4 + (d+1) \, c_5 \big) R \big] \varphi + \big(c_1 + c_2 + (d+1) \, c_3 \big) R^{\mu\nu} \varphi_{\mu\nu}  = (d-1) \big( \square \varphi - D^\mu \varphi_\mu \big) \;,
\end{equation}
while $D^\mu D^\nu E_{\mu\nu} = 0$ can be written in the form:
\begin{align}
\label{DDEmunu}
-m^2( \square \varphi - D^\mu \varphi_\mu) &= iq (1+c_6 ) F^{\rho\alpha} D_\rho \varphi_\alpha + (c_1-2 ) R^{\mu\alpha\nu\beta} D_\mu D_\nu \varphi_{\alpha\beta} + c_2 R^{\mu\alpha} D_\mu \varphi_\alpha \nn \\
&\quad + (1+c_3 e^{i \theta}) R^{\mu\alpha} D_\mu D_\alpha \varphi + c_3 e^{-i \theta} R^{\alpha\beta} \square \varphi_{\alpha\beta} + c_4 R D^\rho \varphi_\rho + c_5 R \square \varphi \nn \\
&\quad + \ldots
\end{align}
where the ellipsis denotes terms that involve at most single derivatives of $\varphi_{\mu\nu}$.

Since there are seven independent terms on the right hand side of \eqref{DDEmunu} and only six coefficients $c_i$, we cannot make the right hand side vanish and we need to invoke an additional assumption about the metric $g_{\mu\nu}$. If we assume that the metric satisfies the Einstein condition \eqref{EinsteinCond} then eq. \eqref{DDEmunu} becomes
\begin{multline}
\label{DDEmunured}
- m^2( \square \varphi - D^\mu \varphi_\mu) =
i q (1+c_6)F^{\rho\alpha} D_{\rho} \varphi_{\alpha} + (c_1-2) R^{\mu\alpha\nu\beta} D_{\mu}D_{\nu} \varphi_{\alpha\beta} \\
+ \frac{2 \Lambda}{d-1} \Big[ \big( c_2  + (d+1)c_4 \big) D^\mu \varphi_\mu + \big( 1 + 2 c_3 \cos \theta + (d+1) c_5 \big) \square \varphi \Big] + \ldots
\end{multline}
The $c_i$'s can now be chosen so that the right hand side 
is equal to $(m_0^2 - m^2) \left(\square \varphi-D^{\mu}\varphi_{\mu}\right)$ with $m_0^2$  a real constant:
\be
\label{crels}
c_1 = 2 \;, \quad c_6 = -1 \;, \quad
-\big( c_2 + (d+1) c_4 \big) = \big( 1 + 2 c_3 \cos \theta + (d+1) c_5 \big) = \frac{d-1}{2\Lambda} \, (m_0^2 - m^2) \;.
\ee
As suggested by the appearance of $m_0^2$ above, this one parameter family of solutions can be absorbed into a rescaling of the mass.
We are led to the following Lagrangian density:
\begin{equation}
\begin{split}
\label{E:ActionInter}
\mathcal{L} = & - |D_\rho \varphi_{\mu\nu}|^2 + 2|D_\mu \varphi^{\mu\nu}|^2 + |D_\mu \varphi|^2 - \big[ D_\mu \varphi^{*\mu\nu} D_\nu \varphi + \text{c.c.} \big] - m^2 \big( |\varphi_{\mu\nu}|^2 - |\varphi|^2 \big) \\
& \quad +2  R_{\mu\nu\rho\lambda} \varphi^{*\mu\rho} \varphi^{\nu\lambda}
- R_{\mu\nu} \varphi^{*\mu\lambda} {\varphi^\nu}_\lambda
- \frac{1}{d+1} R | \varphi |^2
 - i  q F_{\mu\nu} \varphi^{*\mu\lambda} \varphi^\nu_\lambda \;.
\end{split}
\end{equation}
With the addition of a Maxwell term, (\ref{E:ActionInter}) becomes (\ref{E:ActionSimp}).

Before we proceed with the last two constraints,  it will be useful to write down the equations of motion, their trace, their first divergence and their second divergence after \eqref{crels} have been implemented.
The equations of motion are
\bea
E_{\mu\nu} &= (\square - m^2) \varphi_{\mu\nu} - 2 D_{(\mu} \varphi_{\nu)} + D_{(\mu} D_{\nu)} \varphi - g_{\mu\nu} \big[ (\square - m^2) \varphi - D^\rho \varphi_\rho \big] \\
&\quad + 2 R_{\mu\rho\nu\lambda} \varphi^{\rho\lambda} - g_{\mu\nu} \frac {R}{d+1} \varphi - i\frac q2 \big( F_{\mu\rho} \varphi^\rho_\nu + F_{\nu\rho} \varphi^\rho_\mu \big) = 0 \;.
\label{finalEOMs}
\eea
The trace equation (\ref{trEcurved}) reduces to
\be
\label{trEsimp}
E_\mu^\mu = \left[ d \, m^2 - \frac{d-1}{d+1} R \right] \varphi - (d-1) \big( \square \varphi - D_\mu \varphi^\mu \big) = 0 \;.
\ee
The first divergence can be written as
\begin{multline}
\label{DEmu}
D^\mu E_{\mu\nu} = - m^2 (\varphi_\nu - D_\nu \varphi) + iq \left[ \frac32 F^{\mu\rho} D_\mu \varphi_{\rho\nu} - \frac32 F_{\nu\mu} \varphi^\mu + \frac12 F^\mu \varphi_{\mu\nu} \right.\\
\left. - \frac12 F_\nu \varphi + \frac32 F_{\nu\mu} D^\mu \varphi + \frac12 (D_\mu F_{\rho\nu}) \varphi^{\mu\rho} \right] = 0 \;,
\end{multline}
where we have employed the notation $F_\nu \equiv D^\mu F_{\mu\nu}$.
The second divergence  can be written as
\begin{multline}
\label{DDEsimp}
D^\nu D^\mu E_{\mu\nu} = - m^2 (D^\nu \varphi_\nu - \square \varphi) + iq \Big[  (D^\mu F^{\nu\rho}) (D_\nu \varphi_{\rho\mu}) - F^\mu \varphi_\mu + F^\mu D_\mu \varphi \Big] \\
- \frac32\, q^2 F^{\mu\rho} F_{\nu \rho} \varphi_\mu^\nu + \frac34\, q^2 F_{\mu\nu} F^{\mu\nu} \varphi
=0 \;.
\end{multline}

Our penultimate constraint is obtained by combining the second divergence (\ref{DDEsimp}) with the trace equation \eqref{trEsimp}:
\begin{multline}
\label{E:penultimate}
-m^2 \Big[ \frac{d \, m^2}{d-1} - \frac{R}{d+1}  \Big] \, \varphi =
iq \Big[  (D^\mu F^{\nu\rho}) (D_\nu \varphi_{\rho\mu}) - F^\mu \varphi_\mu + F^\mu D_\mu \varphi \Big] \\
- \frac32\, q^2 \Big[ F^{\mu\rho} F_{\nu \rho} \varphi_\mu^\nu - \frac12 \, F_{\mu\nu} F^{\mu\nu} \varphi
\Big] \;.
\end{multline}
Unlike the $q=0$ case, the constraint \eqref{E:penultimate} is not a purely algebraic
constraint on $\varphi$. Instead, it  involves first derivatives of $\varphi_{\mu\nu}$ multiplied by first derivatives of $F_{\mu\nu}$. Thus, it is
not immediately clear that taking a time derivative of this constraint equation will yield
the final constraint. Acting on \eqref{E:penultimate} with $D^t$ we find
\be
\label{lastconstraint}
-m^2 \Big[ \frac{d \, m^2}{d-1} - \frac{R}{d+1}  \Big] \, D^t \varphi =
i q \Big[
-(D_j F^{ji}) D^t D_t \varphi^t_i + (D_i F^{tj}) D^t D_t ( \varphi^i_j - \delta^i_j \varphi^k_k ) + \ldots
\Big] \;,
\ee
where here and below the ellipsis denotes terms that do not involve second order time derivatives of $\varphi_{\mu\nu}$.
We now use the equations of motion (\ref{finalEOMs}) and its divergence (\ref{DEmu}) to eliminate these second order time derivatives. The space-space components of the EOM's give
\begin{equation}
0 = E^i_j = D^t D_t ( \varphi^i_j - \delta^i_j \varphi^k_k) + \ldots \;,
\label{E:EijEOM}
\end{equation}
which we can use to eliminate the second term on the right hand side of (\ref{lastconstraint}). Taking the time derivative of the divergence of the EOM's we get:
\begin{equation}
\label{DtDE}
0 = D^t D_\mu E^\mu_j = - m^2 D^t D_t \varphi^t_j +  \frac{3iq}{2} \big[ {F^i}_j D^t D_t \varphi^t_i
+ {F^t}_i D^t D_t ( \varphi^i_j - \delta^i_j \varphi_k^k ) \big] + \ldots
\end{equation}
Using again \eqref{E:EijEOM}, we can eliminate the second term in the square brackets.
Finally, for generic values of $F_{ij}$ we can invert the matrix
\begin{equation}
m^2 \left[ \delta^i_j - \frac{3iq}{2m^2} {F^i}_j \right]
\end{equation}
and solve (\ref{DtDE}) for $D^t D_t \varphi^t_j$ in terms of first order time derivatives of $\varphi_{\mu\nu}$.  With this solution in hand, we can eliminate also the first term on the right hand side of (\ref{lastconstraint}), and reduce it to the last constraint equation.
As pointed out in \cite{Kobayashi:1978mv}, there will be choices of $F_{ij}$ for which this matrix is not invertible, so that we fail to find a constraint. However for $q{F^i}_j/m^2 \ll 1$, which is one of the requirements in (\ref{cutoff}), the matrix is invertible.

Given the possible non-existence of the last constraint, one wonders if our Lagrangian may have other pathologies.  Indeed \cite{Velo:1970ur,Velo:1972rt} discovered that for generic values of $F_{\mu\nu}$, the EOM's are either non-hyperbolic or non-causal. Failure of hyperbolicity (which is the condition to have a well-posed Cauchy problem) is associated with electric fields, and it is avoided by taking ${F^i}_j$ smaller than a certain bound. On the other hand, non-causal behavior is associated with both electric and magnetic fields, and appears for arbitrarily small values of such fields.  Fortunately this effect appears at an order specified by (\ref{cutoff}) and is small when the field strength is small.
In a frame where ${F^i}_j$ is small this pathology can presumably be corrected by adding terms to the Lagrangian that are higher order in $F_{\mu\nu}$. Such an expansion was carried out in a different context in \cite{Zinoviev:2009hu}.

\section{Stability bound for higher spin fields in AdS}
\label{app:BF bound}

Breitenlohner and Freedman found that there is a lower bound on the mass squared of a free scalar field propagating in AdS${}_{d+1}$ in order for it to be stable and non-tachyonic  \cite{Breitenlohner:1982bm}:
\begin{equation}
\label{E:BFbound}
m^2 L^2 \geq -\frac{d^2}{4} \;.
\end{equation}
Their argument is based on the requirement that the energy functional for the scalar be positive definite.
For fermionic fields, reality of the action requires $m$ to be real so that $m^2\geq 0$.
We find that, for bosonic fields of spin one and higher, the stability bound on the mass squared is also
expected to be
\be
\label{E:BFHigherSpin}
m^2 \geq 0 \;.
\ee
In what follows, we illustrate \eqref{E:BFHigherSpin} for the simple case of a massive vector field.

Consider a massive vector field, whose action is given by
\be
S = \frac12 \int d^{d+1}x\, \sqrt{-g}\, \Big( - \frac12 F_{\mu\nu} F^{\mu\nu} - m^2 A_\mu A^\mu \Big) \;,
\ee
propagating in global AdS$_{d+1}$ with line element
\begin{equation}
ds^2 = (-dt^2 + d\rho^2 + \sinh^2\rho\, d\Omega_{d-1} )/\cosh^2\rho \;.
\end{equation}
The energy functional
\begin{equation}
\label{E:EF}
E = \int d^dx\, \sqrt{-g}\, T^{0\mu} \xi_\mu
\end{equation}
for the spin one field can be constructed by integrating the bulk stress-energy tensor
\begin{equation}
T_{\mu\nu} = \frac{2}{\sqrt{-g}} \, \frac{\delta\cL}{\delta g^{\mu\nu}}
\end{equation}
over a spacelike slice 
orthogonal to the timelike Killing vector $\xi^\mu = (1,0,\dots,0)$. Distinguishing the time component ($\mu = 0$) from the spatial ones ($\mu,\nu = i,j$), one finds that the energy functional \eqref{E:EF} takes the form
\be
\label{energy functional vector}
E = \frac12 \int d^dx\, \tanh^{d-1}\rho\, \Big\{ g^{ij} F_{0i} F_{0j} + m^2 A_0^2 + \frac1{2\cos^2\rho}\, F_{ij} F^{ij} + \frac{m^2}{\cos^2\rho}\, A_i A^i \Big\} \;.
\ee
For $m^2 >0$ the energy functional is manifestly positive definite.
For $m^2 = 0$
the energy functional and action are those of a massless vector field, and are both invariant under gauge transformations.
Solutions with $F_{\mu\nu}=0$ have zero energy, and so the energy functional is semi-positive definite. Interpreting the energy as the norm in the Hilbert space of states, one finds that such states are null and therefore not physical.
When $m^2<0$ solutions with $F_{\mu\nu}=0$ can have negative energy.
The kinetic energy of these solutions is zero and the (negative) mass term in the energy functional is responsible for the energy being negative.%
\footnote{One might be worried that the constraint equation $D_{\mu} A^{\mu} = 0$ prohibits solutions of the form $F_{\mu\nu}=0$. However,
since (\ref{energy functional vector}) does not depend on $\partial_0 A_0$ nor $\partial_\rho A_\rho$, at a given time we can construct modes whose only non-trivial component is $A_\rho(\rho)$, vanishing at large $\rho$ arbitrarily fast, and with $\partial_0 A_\rho = 0$ and $\partial_0 A_0$ fixed by the constraint.}

The crucial point in the above computation is that the action is the sum of two terms: a kinetic term that is invariant under gauge transformations, and a mass term that is not invariant under gauge transformations and gives rise to a term of definite signature in the energy functional.
Shifting our attention to higher spin bosonic fields, whenever the action is the sum of two terms---a kinetic term and a mass term---with the properties just described,
stability of the energy functional should imply $m^2 \geq 0$.
 Note that bosonic fields of spin larger than two require auxiliary fields; therefore our argument regarding positivity of the energy functional requires improvement.

Positivity of the energy functional is meaningful only when it is finite and can be used to bound $\Delta$.
Considering the neutral spin two field with BGP Lagrangian \cite{Buchbinder:2000fy} (\ref{E:ActionSimp}),
with the addition of suitable boundary terms, the energy functional (or, equivalently, the on-shell action)
converges if $\varphi_{\mu\nu}$ vanishes at large $\rho$ faster than $(\cosh\rho)^{\frac{d-6}2}$.
Since $\varphi_{mn} \sim \cosh^{\Delta-2}\rho$, $\Delta$ should be positive for $d \geq 3$.
Given the relation $m^2 = \Delta (\Delta - d)$ and the bound $m^2 \geq 0$,
we conclude that in fact $\Delta \geq d$. As opposed to the scalar case where the leading and subleading terms of a near-boundary series expansion can change roles \cite{Klebanov:1999tb}, the spin two field has only one possible quantization scheme.

\section{Approximating $T_c$ in the classical limit}
\label{app:semiclassics}

In section \ref{S:dwaveCondensate} we have argued that the system of equations governing the phase transition for a massive charged spin two field is identical to that for  a massive charged scalar field. When the mass of the scalar field becomes very large we can treat the system semi-classically.

Consider a massive, charged particle in a charged AdS black hole background. Such a particle experiences an attractive gravitational force toward the black hole and a repulsive electrostatic force towards the asymptotically AdS boundary. For a high temperature black hole the gravitational force will always dominate over the electrostatic repulsion, and there will be no equilibrium configuration.  However, as the black hole cools, the gravitational force close to the horizon becomes weaker, and one finds that the particle will settle in a global minimum of its potential, away from the black hole horizon. We identify the temperature at which this global minimum appears with the temperature of the phase transition for the scalar condensate.

\begin{figure}
\begin{center}
\includegraphics{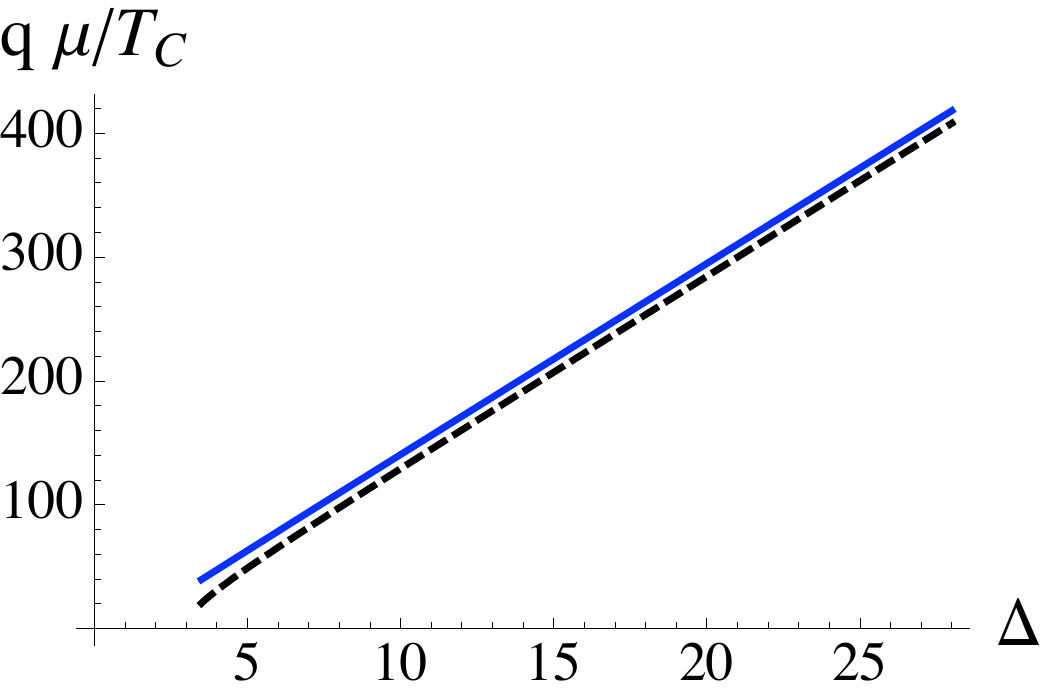}
\caption{\label{F:Tcplot} (Color online) The critical temperature for generating a holographic $d$-wave or $s$-wave condensate as a function of the dimension of the condensing operator. The numerical value for the critical temperature is plotted in solid blue. The approximation \eqref{E:TcAnalytic} is plotted as a dashed black line.}
\end{center}
\end{figure}

The action for a massive, charged point particle is given by
\begin{equation}
\label{E:Spp}
S_{pp} = -m \int d\tau - q \int A_{\mu} dx^{\mu} \;.
\end{equation}
The effective potential for a static configuration in the background \eqref{E:normalphase} and in $d=3$ boundary dimensions is given by
\begin{equation}
\label{E:Vppval}
	V_{pp}(\zeta) = \frac{mL}{z_h}\frac{\sqrt{1-\zeta^3}}{\zeta} + \frac{\zeta-1}{z_h} \tilde{\mu}
\end{equation}
where we have defined $\zeta = z/z_h$ and $\tilde{\mu} = q \mu z_h$.  We obtained \eqref{E:Vppval} by evaluating the point particle action \eqref{E:Spp} on a solution where the point particle is stationary. The minimum of the potential is located at a $\zeta_{min}$ for which
\begin{equation}
\label{E:Vpcondition}
	V_{pp}'(\zeta_{min})= 0 \;.
\end{equation}
Since $V(1)=0$ this minimum will be a global minimum only when $\tilde{\mu}$ is large enough so that
\begin{equation}
\label{E:Vcondition}
	V_{pp}(\zeta_{min}) \leq  0 \;.
\end{equation}
A quick computations shows that \eqref{E:Vcondition} and \eqref{E:Vpcondition} are satisfied whenever $\tilde{\mu} > \tilde{\mu}_c$ where
\begin{equation}
\label{E:TcAnalytic}
	\tilde{\mu}_c = \sqrt{\frac{14}{3} + \frac{1}{12} \left(152072-1224\sqrt{17}\right)^{1/3}+\frac{1}{6}\left(19009+153\sqrt{17}\right)^{1/3}} \, mL
	\sim 3.68 \, mL \,.
\end{equation}
In figure \ref{F:Tcplot} we compare the expression \eqref{E:TcAnalytic} with a numerical evaluation of the normal modes of a condensing scalar field.
Note that the numerical value for $T_c$ is slightly higher than the classical approximation \eqref{E:TcAnalytic}.  This is, perhaps, expected: in the quantum theory the ground state wavefunction has some kinetic energy so the minimum of the potential ought to be slightly lower than zero.

\section{Fermion Green's functions}
\label{app:FGreens}

We define the retarded Green's function for fermions
in analogy to the Green's function for a scalar operator:
\begin{equation}
\label{E:GRfermion}
\widetilde G_R(t, \vec x) = i \Theta(t) \langle \{ \cO_\Psi(t, \vec x), \cO_\Psi^\dag(0) \} \rangle \;.
 \end{equation}

In order to derive the retarded fermionic Green's function from AdS/CFT, we must
first obtain the explicit form of the boundary terms in the fermion action $S_{\rm bnd}$. The variation of the Lagrangian \eqref{E:FermionAction} which leads to the Dirac equation \eqref{Dirac equation} also results in a boundary term of the form
\begin{equation}
\delta S_{\rm bulk} = -i  \int d^3 x ( \overline R \, \delta S - \overline S \, \delta R) \ ,
\end{equation}
where $R$ and $S$ were defined in (\ref{RSdef}) and are three dimensional spinors.
In minimizing the action, we would like to vary $S_\Psi$ such that $\delta S$ vanishes on the boundary while $\delta R$ is unconstrained.  Thus, we need to add to the action a boundary counter term in order to have a well defined variational principle. An appropriate boundary action is
\begin{equation}
S_{\rm bnd} = -i \int d^3x \, \overline S R
= -\frac{i}{2}  \int d^3 x \sqrt{-g_{\rm bnd}} \left( \overline \Psi \Psi + \overline \Psi n_\mu \Gamma^{\mu} \Psi \right) \;,
\end{equation}
where $n_\mu$ is a unit vector normal to the boundary, pointing toward positive $z$.
The total variation of the action is then
\begin{equation}
\delta S_{\Psi} = \delta S_{\rm bulk} + \delta S_{\rm bnd} = -i  \int d^3x \, ( \overline{\delta S} \, R + \overline R \, \delta S)
\label{Svary}
\end{equation}
which is independent of $\delta R$ as required.

Having found the appropriate boundary counter term, we return to a construction of the fermionic Green's function.  Instead of thinking of (\ref{Svary}) as a result of varying the action,
we will now think of it as a perturbation to the action and ask how the fermionic operators
respond at linear order to the presence of a $\delta S \neq 0$.
From the theory of linear response, an expression that describes the change in the expectation value of an operator ${\mathcal O}$ due to the addition of a term $\delta H$ to the Hamiltonian is
\begin{equation}
\delta \langle {\mathcal O}(x) \rangle = - i \int dt' \, \Theta(t-t') \langle [ {\mathcal O}(x), \delta H(t') ] \rangle \;.
\end{equation}
For the fermionic operator $R$, we have $\int dt \, \delta H = - \delta S_{\Psi}$ and
\begin{eqnarray}
\delta \langle R(x) \rangle &=&
 \int d^3x' \Theta(t-t')
\left[
\langle \{ R(x) , \overline R(x') \} \rangle \delta S(x')
+ \langle \{ R(x), \overline{R^c}(x') \} \rangle \delta S^c(x')
\right]
\\
&=&
-i \int d^3x' \, \left[ \widetilde G_R(x-x') \gamma^t  \delta S(x') + \widetilde G_R^c(x-x')  \gamma^t  \delta S^c(x') \right] \;.
\end{eqnarray}
The Fourier transform of this relation is the multiplicative identity
\begin{equation}
R^{(\omega, \vec k)} = -i  \big(  G_R(k) \gamma^t S^{(\omega, \vec k)}+  G_R^c(k)
\gamma^t S^{(-\omega, -\vec k)c} \big) \;,
\label{FTlinrespond}
\end{equation}
where $G_R(k)$ is the Fourier transform of $\widetilde G_R(x)$.
Comparing (\ref{FTlinrespond}) with the bulk computation (\ref{bulklinrespond}), we
deduce the form of the retarded Green's function (\ref{GRanswer}), consistent with the result
of \cite{Iqbal:2009fd}.

Before wrapping up this section, we would like to check that we have chosen the sign of the fermionic action such that the spectral function
$\rho(k) = \Tr \Im G_R(k)$ is positive.
For $k = (\omega, 0)$ and $m_\zeta=0$ the dimension six operator in \eqref{E:FermionAction} vanishes and the Dirac equation can be solved exactly. We find that $\cM(\omega,0) = \gamma^t$ and therefore $\rho(\omega, 0) = 2$.

This result for $\rho(\omega,0)$ suggests the following sum rule
\begin{equation}
\int d \omega \, \big( \rho(\omega, \vec k) - 2 \big) = 0 \;,
\end{equation}
which we have been able to verify numerically for a few generic values of $\vec k$.
Analytically, we have found that for $\eta = 0$, $\rho(\omega, \vec k ) = 2 + O(1/\omega^2)$,
although we believe the result holds more generally.
Moreover, we expect that $G_R$ is analytic in the upper half of the complex $\omega$ plane.  Thus one ought to be able to demonstrate the sum rule through a contour integration.

\bibliography{bib2}{}
\bibliographystyle{utphys}

\end{document}